\documentclass[a4paper,fleqn,usenatbib]{mnras}

\usepackage{newtxtext,newtxmath}
\usepackage[T1]{fontenc}
\usepackage{ae,aecompl}

\usepackage{graphicx}	% Including figure files
\usepackage{amsmath}	% Advanced maths commands
\usepackage{amssymb}	% Extra maths symbols
\usepackage{slashed, epsf, color}
\usepackage{mathrsfs}
\usepackage{url}
\usepackage{comment}
\usepackage{ifthen}

\newcommand{\unit}[1]{\,\text{#1}}
\newcommand{\Order}[1]{{\cal O}(#1)}

\newcommand{\mDM}{m_{\text{DM}}}

\newcommand{\bani}{\beta_{\text{ani}}}

\newcommand{\Besancon}{Besan\c{c}on}

\newcommand{\tMem}{{\text{Mem}}}
\newcommand{\tFG}{{\text{FG}}}
\newcommand{\tBes}{{\text{Bes}}}

\newcommand{\emp}{{\tilde{p}}}
\newcommand{\emv}{{\tilde{v}}}
\newcommand{\ems}{{\tilde{\sigma}}}
\newcommand{\Likeli}{{\mathcal{L}}}
\newcommand{\emLikeli}{\tilde{\mathcal{L}}}
\newcommand{\emfmem}{{f_\tMem}}
\newcommand{\emffg}{{f_\tFG}}
\newcommand{\emvmem}{{\emv_\tMem}}
\newcommand{\emsmem}{{\ems_\tMem}}

\newcommand{\emsfg}{{\ems_\tFG}}
\title[Foreground effect on the $J$-factor estimation]{
Foreground effect on the $J$-factor estimation\\
 of classical dwarf spheroidal galaxies
}

\author[K. Ichikawa et al.]{
Koji Ichikawa,$^{(a)}$
Miho N. Ishigaki,$^{(a)}$
Shigeki Matsumoto,$^{(a)}$
Masahiro Ibe,$^{(a, b)}$\newauthor
\ Hajime Sugai,$^{(a)}$
Kohei Hayashi,$^{(a, c)}$
and Shun-ichi Horigome$^{(a)}$
\\
$^{(a)}${Kavli IPMU (WPI), UTIAS, The University of Tokyo, Kashiwa, 277-8583, Japan} \\
$^{(b)}${ICRR, The University of Tokyo, Kashiwa, 277-8582, Japan} \\
$^{(c)}${Kavli Institute for Astronomy and Astrohysics, Peking University, Beijing 100871, China}
}

\date{Accepted XXX. Received YYY; in original form ZZZ}

\pubyear{2016}

\begin{document}
\label{firstpage}
\pagerange{\pageref{firstpage}--\pageref{lastpage}}
\maketitle

\begin{abstract}
The gamma-ray observation of the dwarf spheroidal galaxies (dSphs) is a promising approach to search for the dark matter annihilation (or decay) signal.
The dSphs are the nearby satellite galaxies with a clean environment and dense dark matter halo so that they give stringent constraints on the $\Order{1} \unit{TeV}$ dark matter.
However, recent studies have revealed that current estimation of astrophysical factors relevant for the dark matter searches
are not conservative, where the various non-negligible systematic uncertainties are not taken into account.
Among them, the effect of foreground stars on the astrophysical factors has not been paid much attention, which becomes more important for  deeper and wider stellar surveys in the future. 
In this article, we assess the effects of the foreground contamination by generating the mock samples of stars and using a model of future spectrographs.
We investigate various data cuts to optimize the quality of the data and find that the cuts on the velocity and surface gravity can efficiently eliminate the contamination.
We also propose a new likelihood function which includes the foreground distribution function.
We apply this likelihood function to the fit of the three types of the mock data (Ursa\,Minor, Draco with large dark matter halo, and Draco with small halo) and three cases of the observation.
The likelihood successfully reproduces the input $J$-factor value while the fit without considering the foreground distribution gives large deviation from the input value by a factor of three.
\end{abstract}

\begin{keywords}
galaxies: dwarf  spheroidals-- galaxies:kinematics and dynamics -- $\gamma$-rays: galaxies-- instrumentation: spectrographs--dark matter --astroparticle physics
\end{keywords}

%%%%%%%%%%%%%%%%%%%%%%%%%%%%%%%%%%%%%%%%%%%%%%%%%%

%%%%%%%%%%%%%%%%% BODY OF PAPER %%%%%%%%%%%%%%%%%%

%------------------------------------------------------------------------------
%---------------------          Introduction          -------------------------
%------------------------------------------------------------------------------
\section{Introduction}
\label{sec:intro}
The existence of  dark matter is solidly confirmed by various observations such as the dynamics of galaxy clusters\,\citep{1933AcHPh...6..110Z}, rotation curves of the galaxies\,\citep*{1978ApJ...225L.107R,1980ApJ...238..471R} and gravitational lensing\,\citep{McLaughlin:1998sb, Lokas:2003ks,Clowe:2006eq,Bradac:2006er} as well as the global fit of the Cosmic Microwave Background (CMB), Large Scale Structure (LSS), and Supernovae (SNs) observations\,\citep{Ade:2015xua}.
Its identity, however, has remained unknown for almost eighty years since its first postulation. 
The identification of dark matter is certainly one of the most important tasks in cosmology, astrophysics, and particle physics.
%%%%%%%%%%%%%%%%%%%%
Despite our limited knowledge of its nature, 
it is  strongly expected that dark matter is not a part of 
the standard model of the elementary particle physics.
Among various candidates for dark matter, weakly interacting massive particle (WIMP) is considered to be
one of the most attractive candidates. 
The thermally produced WIMP dark matter can explain the observed dark matter density naturally 
by its annihilation into lighter particles in the standard model.
In fact, the preferred mass range of the WIMP is $\lesssim \Order{1}$\,TeV since otherwise its annihilation rate is suppressed 
and the mass density becomes inconsistent with the observation.
With this successful feature,  the WIMP dark matter has gathered particular attention  in conjunction
with the physics beyond the standard model such as supersymmetry (see e.g. \citet*{Jungman:1995df} also \citet{Murayama:2007ek,Feng:2010gw}).
%%%%%%%%%%%%%%%%%%%%
%There are three major methods for WIMP dark matter detections: collider searches, direct detections, and indirect detections.
%Collider experiments may directly produce the dark matter when its centre-of-mass energy is sufficiently above the dark matter mass\,\citep{Tsukamoto:1993gt}, where the produced dark matter is seen as missing energy.
%Direct detection experiments search for the scattering process between the dark matter and nuclei. 
%Recent experiments with two-layer xenon system such as XENON-100\,\citep{Aprile:2012nq}, LUX\,\citep{Akerib:2013tjd} 
%have strong sensitivity for $\Order{10} \unit{GeV}$. 
%However, it is still challenging for the collider experiments and the direct detection experiments
%to search for the most interesting region of the WIMP dark matter, i.e., $\Order{1} \unit{TeV}$ mass region.

Indirect detection experiment, which aims to observe the signals induced by the dark matter annihilation, is one of the most important searches for the WIMP dark matter
because it has the strong sensitivity to the most interesting mass scale, i.e., $\Order{1} \unit{TeV}$.
%are \note{a complementary approach} to search 
In particular, the gamma-ray search is a promising approach because it is free from the propagation uncertainty and the detection capability is sufficiently large. 
Many astronomical objects potentially are to be targets of the gamma-ray dark matter detection.
Among them, dwarf spheroidal satellite galaxies of the Milky Way (dSphs) are ideal targets because they are nearby galaxies ($\sim 10\--100 \unit{kpc}$ from the solar system) at high latitude with dense dark matter halo.\footnote{
The observation of the galactic centre of the Milky Way is also sensitive to the dark matter signal. 
However, the signal flux highly depends on the shape of the dark matter profile and 
understanding of the astrophysical background is difficult.
The most conservative estimation is weaker than that of dSphs.} 
In fact, current observations have given stringent constraints on the dark matter annihilation cross sections
by searching for the gamma-ray signal of dark matter\,\citep{Ackermann:2015zua} from dSphs.
As pointed out in \citet{Bhattacherjee:2014dya}, however, 
the constraints crucially depend on the astrophysical uncertainties of the dark halo shape of the 
dSphs especially the ones of the so-called ultra-faint dSphs whose numbers of the observed stars are less than $\Order{100}$.
Recent detailed analyses have also revealed hidden but non-negligible uncertainties of the dark matter profile: 
prior bias\,\citep{Martinez:2009jh}, anisotropy of the velocity dispersion\,\citep{Ullio:2016kvy},
size of the halo truncation\,\citep{Geringer-Sameth:2014yza}, 
non-sphericity\,\citep{Bonnivard:2015xpq, Hayashi:2016kcy}, and foreground contamination\,\citep{Bonnivard:2015vua}.\footnote{
The existence of the unresolved binary stars can also have impact on the estimation of the velocity dispersion.
For classical dSphs, 
although the typical fractions lie around $20\--60$\,\%,
the effect of these binary stars is negligible
because their intrinsic velocity dispersions are much larger than those of the binary stars
\,(\citealp{Olszewski:1995gs, Walker:2005nt}; \citealp*{Mateo:2007xh}; \citealp{Koch:2007ye, Minor:2013bj}).
For some ultra-faint dSphs,
the velocity variability due to the unresolved binary stars is measured and 
it is confirmed that the binary effect does not significantly inflate the velocity dispersion
\,\citep{Simon:2007dq,Simon:2010ek,McConnachie:2010pn,Koposov:2011zi,Kirby:2013isa,Simon:2015fdw}.
However, one should make clear this effect when one use dataset of a new ultra-faint dSph,
which will be resolved by the multi-epoch velocity data.
}

These systematic uncertainties are partly caused by the lack of sufficiently 
large stellar kinematic data available from current spectroscopic instruments. 
For the eight luminous dSphs (so-called classical dSphs), 
discovered before the Sloan Digital Sky Survey project\,\citep{York:2000gk}, 
\citet{0906.0341} used line-of-sight velocity measurements of up to a few thousand stars in each galaxy to constrain their dark matter content, based on the observations with the Michigan/MIKE Fibre System at the Magellan 6.5 m Clay telescope and the Hectochelle spectrograph at the 6.5 m Multiple Mirror Telescope. The number of stars is limited by their limiting magnitudes ($i$-band magnitudes of up to $\sim 19.5$) and/or the small field-of-view compared to the tidal radii of these galaxies. 
For the ultra-faint dSphs, the number of stars is intrinsically small and thus even with a multi-object spectrograph mounted on a $8\--10$ m class telescope (e.g. Deep Imaging Multi-Object
Spectrograph at Keck telescope), 
the velocity measurements for only a few tens to a few hundred stars are available\,\citep{
Simon:2007dq,
Simon:2010ek,
Kirby:2013isa,
Koposov:2011zi, 
Aden:2009dz, 
Walker:2009iq,
Simon:2015fdw,
2015ApJ...808...95S,
2016ApJ...818...40M,
Kirby:2015ija}.  
However, since most of the member stars exist below their sensitivity ($i \sim 21$), 
the number of the measured velocities  is about $\Order{10\--100}$.

Most of the uncertainties mentioned above would be reduced with future wide-field multi-object spectrographs mounted on large telescopes that are capable of efficiently covering the outer regions of dSphs and reaching fainter magnitudes (e.g. $i>21.0$).
However, the spectroscopic data always includes
the foreground stars which belong to the Milky Way galaxy.
Such contamination will remain problematic 
or even become worse in the future observations. 
Current approach to eliminate the foreground stars is to calculate 
their membership probability, 
which is obtained through the expectation maximization process using radial velocity, 
projected position, and metallicities\,\citep{Walker:2008fc}.
However, once the stars with high membership probability are chosen, 
they are usually considered as the true member stars
and the effect of the residual foreground stars 
is ignored in the halo profile estimation.
As \citet{Bonnivard:2015vua} have pointed out,
this contamination can derive a non-negligible overestimation of $J$-factor and 
therefore more detailed analysis of the foreground contamination is required for the future detailed kinematical survey.

In this paper, we investigate the foreground effect in the future observation.
The data cuts which minimize the fraction of the foreground stars are investigated 
by using mock data under a realistic setup of the future spectrographic observation, 
which provides the prospect of the number and quality of the future stellar data at the same time.
We also construct a new method for the $J$-factor estimation convolving the foreground contamination.
This method is compared with the usual halo estimation which gives largely biased $J$-factor by the foreground contamination.
The organization of this paper is as follows.
In section 2, we introduce a formula of the signal flux and clarify the uncertainties which are not taken into accounted in the current analysis.
In section 3, we construct mock data of dSph member stars and foreground stars.
The number of expected observed stars are estimated by assuming the model of the spectrograph and imposing the data cut.
After the cut, we give a new likelihood function including the foreground contamination.
The results of the fit using the new likelihood are given in section 4.
Finally, we summarize our discussion in section 5.
In appendix A, we give 
the detailed method to estimate the foreground distribution required for the new likelihood function.
We compare our results with those obtained by a conventional analysis in appendix B.
Appendix C is devoted to show the dependence of the halo truncation radius on the $J$-factors.

%------------------------------------------------------------------------------
%-------------------------          Flux          -----------------------------
%------------------------------------------------------------------------------
\section{Gamma-ray flux formula}
The differential gamma-ray flux from the dark matter annihilations in a solid angle $\Delta \Omega$ is given by
\begin{eqnarray}
\Phi (E, \Delta \Omega)
=
\left[
\frac{C \langle \sigma v \rangle}{4 \pi \mDM^2}
\sum_{f}
b_{f}
\left(\frac{dN_\gamma}{dE}\right)_f
\right]
\left[
\rule{0ex}{4ex}
\int_{\Delta \Omega} d\Omega \int_{l.o.s.} dl \,
\rho^2(l, \Omega)
\right]\ .
\label{eq:fluxformula}
\end{eqnarray}
Here $C$ is 1/2 for Majorana and 1/4 for Dirac dark matter,
$\mDM$ denotes the mass of the dark matter,
and $\langle \sigma v \rangle$ represents the product of the total annihilation cross section and the relative velocity $v$ that is averaged with the velocity distribution function. 
The branching fraction of the annihilation into a final-state $f$ is denoted by $b_f$, and $(dN_\gamma/dE)_f$ is the differential number density of photons for a given final state $f$.
The dark matter profile inside dSph is defined by $\rho(l, \Omega)$.
The integration of the profile is done along the line-of-sight in the region of interest (ROI) $\Delta \Omega$.

We here note the uncertainties coming from the first parenthesis in the right-hand side of Eq.(\ref{eq:fluxformula}).
This part is determined only by particle physics.
Usually, the (velocity-averaged) total cross section and branching fraction can be obtained by a perturbative calculation and the error can be reduced to $\lesssim \Order{1}$ percent by a higher order calculation.
The fragmentation function $(dN_\gamma/dE)_f$ usually gives larger errors.
For examples, the annihilation products might be fragmented into various hadrons,\footnote{
The annihilation can also produce monochromatic gamma-rays directly via, for example, $Z \gamma$, $\gamma \gamma$ channel.}
which eventually produces stable particles (such as $p$, $\bar{p}$, $e^{\pm}$, $\gamma$, and $\nu$).
The energy distribution of the photons is calculated by a Monte Carlo simulation
such as Pythia\,\citep*{Sjostrand:2007gs} or HERWIG\,\citep{Corcella:2000bw}, which includes the effect of QED and QCD final-state radiations.
The errors become large in the region of small photon energy (especially for $\tau^{+}\tau^{-}$ channel)\,\citep{Cirelli:2010xx,Cembranos:2013cfa}.
However, in most cases, the uncertainty is $\lesssim 10\,\%$ as long as the gamma-ray is hard enough (above $\sim 0.1 \,\%$ of the dark matter mass).
Because the current gamma-ray detectors are sensitive for ${\cal O}$(0.1--1000)\,GeV gamma-ray,
the uncertainties in the particle factor do not significantly alter the sensitivities for the ${\cal O}$(1)\,TeV dark matter.
Below, we discuss more important uncertainties of the astrophysical factor in the second parenthesis, 
where the error can reach two orders of magnitude.

%------------------------------------------------------------------------------
\subsection{Astrophysical factor}
\label{subsec:systematics}
The second parenthesis in the right-hand side of Eq.\,(\ref{eq:fluxformula}), so called the $J$-factor, represents the amount of the dark matter in the halo deduced from astrophysical observation.
Using the kinematical data obtained from the spectroscopic surveys, many studies provide the estimations of the $J$-factors.
In \citet{Martinez:2013els}, the dark matter profile is evaluated by using the Bayesian hierarchical modelling where they fit the halo parameter by assuming empirical relations between the maximum velocity, maximum radius, and total luminosity with several free parameters.
The fit is performed by imposing these relations as priors, while the free parameters of the relations are simultaneously optimized by utilizing all the data of multiple dwarf galaxies.
On the other hand, one can also directly estimate the dark matter halo by comparing the stellar velocity data with the theoretical dispersion curve.
\footnote{
Here we note that a spectrograph can only measure stellar velocities along the line-of-sight, 
and they cannot be directly used for the halo estimation.
}
The dispersion curve is  obtained from the Jeans equation\,\citep{2008gady.book.....B} under the assumption of spherical symmetry and steady (and dark matter dominated) system, which is expressed as
\begin{eqnarray}
\frac{1}{\nu_{*}(r)}
\frac{\partial}{\partial r}
\nu_{*}(r) \sigma^{2}_{r}(r)
+
\frac{2 \beta_{\text{ani}} (r) \sigma^{2}_{r}  (r)}{r}
=
- \frac{G M(r)}{r^2}\ ,
\label{eq:Jeans1}
\end{eqnarray}
where $r$ denotes the distance from the centre of the dSph and 
$\nu_{*}(r)$ is the number distribution of the dSph member stars 
obtained from photometric observations.
The velocity dispersions of the stars in the dSph are defined by
$\sigma_{r}$, $\sigma_{\theta}$, and $\sigma_{\phi}$, which denote the components 
along the radial, azimuthal, and  polar direction respectively. 
Here, for the spherical symmetry, we take $\sigma_{\theta}  =  \sigma_{\phi}$.
The anisotropy parameter $\beta_{\text{ani}}$ is defined by
$\beta_{\text{ani}} = 1 - \sigma^{2}_{\theta}/\sigma^{2}_{r}$, $G$ is the gravitational constant, and $M(r)$ is the enclosed mass of the dark matter halo.

To compare this velocity dispersion with the observables, one should project 
it along the line-of-sight.
A straightforward calculation gives the projected dispersion curve $\sigma_{l.o.s}$:
\begin{eqnarray}
\sigma^{2}_{l.o.s}(R) = \frac{2}{\Sigma_{*}(R)} 
\int^{\infty}_{R}  dr
\left( 1 - \beta_{\text{ani}} (r) \frac{R^2}{r^2} \right)
\frac{\nu_{*} (r) \sigma^{2}_{r} (r)}{\sqrt{1 - R^2/r^2}}
\ ,
\label{eq:dispersion}
\end{eqnarray}
where $R$ denotes the projected distance from the centre of the dSph and $\Sigma_{*}(R)$ is the projected stellar distribution obtained by integrating $\nu_{*} (r) $ along the projected direction.
The fit is performed by comparing the observed $\sigma_{l.o.s}$ and the right-hand values of equation~(\ref{eq:dispersion}) calculated with several fitting parameters (typically the halo profile parameter and velocity anisotropy), with respect to $R$.

Recently, however, it is pointed out that there exist non-negligible systematic errors hidden in the halo estimation: the prior biases, velocity anisotropy, halo truncation, non-sphericity, and foreground contamination.

The prior biases are required for the fit with small observational data.
When the number of the observed stars is small ($\lesssim {\cal O}(100)$), the $J$-factor obtained from the fit does not converge well and gives a large uncertainty (by two orders of magnitude or more)\,\citep*{Bonnivard:2015vua}.
Therefore, most studies evaluate the $J$-factor by imposing  prior biases\,\citep{Martinez:2013els}.
However, \citet{Martinez:2009jh} reveals that the choice of the prior 
strongly affects the halo estimation by at least two orders of magnitude for ultra-faint dSph fit.
\footnote{
Another approach is to impose the empirical parameter cut\,\citep{Bonnivard:2015xpq}.
However, the physical interpretation of this cut is still unclear.
}
Thus, conservatively speaking,
one should be careful when considering the gamma-ray sensitivity lines including the contribution from the ultra-faint dSph.
The effect of the prior biases becomes small when the number of the observed stars is $\sim {\cal O}(1000)$,
and therefore a future stellar observation of the ultra-faint dSphs is essential.

The velocity anisotropy $\bani (r)$ is another subject of the discussion\,\citep{Ullio:2016kvy}. Because the anisotropy parameter cannot be directly addressed, one should make assumptions on the spatial dependence of the anisotropy. Currently, most studies assume that the anisotropy is $r$-independent, while the recent study\,\citep{Bonnivard:2015xpq} fits the kinematical data by using the Baes \& van Hase parametrization\,\citep{Baes:2007tx}. Since it is pointed out that the anisotropy parameter might give a non-negligible effect on the $J$-factor estimation\,\citep{Ullio:2016kvy}, more quantitative discussion should be given in the future.

A further systematic error comes from the morphology of the outer halo.
Even for the classical dSphs with $\sim 500$ member stars, the fit often allows quite large dark matter halo radius (even more than 100 kpc) with small dark matter density\,\citep*{Geringer-Sameth:2014yza} because the star kinematics does not provide the information over the outermost star.
Although the dark matter halo may be truncated at some distance by the effect of the tidal stripping,
there is no consensus on the truncation radius.
One can calculate the tidal radius of the dSphs by assuming the Milky Way halo mass and profiles.
Another conservative approach is to consider the distance of the outermost star as a lower bound of the truncation radius.
Because smaller truncation radius gives smaller $J$-factor, this lower bound method always provides conservative results.
The radius of the outermost star can be estimated by using the projected radius of the stars\,\citep{Geringer-Sameth:2014yza}.
In our analysis, we set the truncation radii based on this estimation
and to eliminate the fluctuation due to the halo truncation, 
we fix the size of the halo truncation at $2000\,\text{pc}$,
which is slightly larger truncation radius than those given by 
\citet{Geringer-Sameth:2014yza} (1.9 kpc for Draco and 1.6 kpc for Ursa\,Minor).
%\footnote{
%We set slightly larger truncation radii than those given by \citet{Geringer-Sameth:2014yza} (1.9 kpc for Draco and 1.6 kpc for Ursa\,Minor).
%}
%In our analysis, we adopt the truncation radius obtained by this estimation.
This estimation will become more accurate when larger stellar data are obtained in the future.

Recent studies also test the contribution from the non-sphericity.
Although most calculations assume the spherical dark matter halo for simplicity, there is no reason why the dark matter halo should be completely spherical.
Thanks to the recent improvement of the numerical resolution, N-body simulation can investigate the shape, orientation, and alignment of dSphs\,\citep{Jing:2002np,Vera-Ciro:2014ita}.
They show that the axis ratio is $\sim 0.6$ for most of the subhalos.
Following this motivation, \citet{Hayashi:2016kcy} performs the axisymmetric fit to the observed stellar data.
The results give $60\--100 \,\%$ deviation for the $J$-factors from the spherical estimation.
A consistent result is also derived by \citet{Bonnivard:2015xpq},
where they estimate the effect of the non-sphericity using mock stellar samples.

The uncertainties above can be improved as the number of the stars increases.
Especially, the uncertainty of the ultra-faint dSphs is dominated by the statistical error (and hence the prior bias) and having more stars significantly suppresses their uncertainties.
Meanwhile, the uncertainty of the classical dSphs stems from the choice of the model of the anisotropy and DM profile, which requires larger amount of data.
Nevertheless, we expect that these model ambiguities will be resolved 
since the stellar data of the classical dSphs can be $3\--4$ times larger  in the future due to the rich stellar population.
In addition, future long-time observation will provide the proper motion which help determining the models.
%It is difficult to significantly reduce systematic errors above with the current kinematical data.
%The simplest way to resolve them is to increase the number of the observed stars and 

Although the future spectroscopic survey toward the dSphs is highly motivated from the points of view of the dark matter detection,
the foreground contamination remains problematic because the number of the foreground stars also increases with the deeper and wider surveys.
%Finally, foreground contamination can also be a non-negligible systematics for the $J$-factor estimation.
%which has non-negligible impact both for the ultra-faint and classical $J$-factor estimation.
%Finally, we introduce the hidden systematics from the foreground contamination.
The observed data always include the stars belonging to the Milky Way galaxy.
Among them, dSph member stars are identified or weighted by membership probability\,\citep{Walker:2008fc} by utilizing the star information such as its position, velocity, colour, metallicity, effective temperature and surface gravity.
However, it is still difficult to absolutely eliminate the foreground contamination.
For instance, \citet{Bonnivard:2015vua} reveals that the profile estimation of the ultra-faint dSph (Segue\,I) is significantly affected by the `marginal' stars, which cannot be clearly identified as member star.
In the Segue\,I case, the overestimation of the $J$-factor reaches more than two orders of magnitude.
The study also shows that the overestimation generally occurs when the number of the observed stars is $\lesssim {\cal O}(100)$.
Moreover, in the membership probability calculation, velocity dispersion of the member stars is assumed to be constant for its position. 
The velocity dispersion of dSph usually changes by factor $1.5 \-- 2$ for its position and therefore this constant assumption might become a non-negligible bias for the membership selection process and subsequent dark matter profile estimation.
The halo estimation including both the foreground contamination and 
the position dependence of the radial velocity dispersion is required for the detailed kinematical analysis
of dSphs.

The next section is devoted to investigate the potential of the future spectroscopic survey as well as 
the reduction of this foreground effect by introducing a new likelihood.

%------------------------------------------------------------------------------
%-----------------------          Analysis          ---------------------------
%------------------------------------------------------------------------------
\section{Analysis}
%------------------------------------------------------------------------------
In the analysis, we first generate realistic mock dSph stellar data including the foreground stars.
To clarify the effect of the contamination, the mock dSph stars are generated assuming the spherical distribution and a constant velocity anisotropy.
Using this mock data, we test the capability of the future spectrograph 
and discuss the efficient data cuts.
Finally, we propose a new likelihood function to eliminate the foreground bias efficiently and fit the mock data using the likelihood function.
Here, we note that the fit is performed under the same assumption above (spherical, constant anisotropy).

\subsection{Dark Matter halo and stellar distribution}
In this paper, we adopt the generalized dark matter halo density profile\,\citep{Hernquist:1990be,Dehnen:1993uh,Zhao:1995cp} as the input dark matter profile for the mock data and fit:
\begin{eqnarray}
\rho_{\text{DM}}(r) = \rho_s (r/r_s)^{-\gamma} (1 + (r/r_s)^{\alpha})^{-(\beta-\gamma)/\alpha}\ ,
\label{eq:DMprofile}
\end{eqnarray}
where $r$ denotes the distance from the centre of the dSph,
 and parameters $\rho_s,\,r_s$ represent the typical density and scale of the halo respectively,
while parameters $\alpha,\,\beta,\,\gamma$ determine the shape of the halo density profile.
For instance, $(\alpha, \beta, \gamma) = (1,\,3,\,1)$ represents the NFW profile\,\citep*{Navarro:1996gj} and $(\alpha, \beta, \gamma) \sim (1.5,\,3,\,0)$ is the Burkert profile\,\citep{Burkert:1995yz,Salucci:2000ps}.
We also assume Plummer profile\,\citep{1911MNRAS..71..460P} for the member stellar distribution:
\begin{eqnarray}
\nu_{*}(r) = (3/4 \pi r_{e}^{3})\,(1 + (r/r_{e})^{2})^{-5/2} \ .
\label{eq:stellardist}
\end{eqnarray}
Here $r_{e}$ denotes the projected half-light radius of the dSph. We normalize $\nu_{*}(r)$ to $\int 4 \pi r^2 \nu_{*}(r) dr = 1$.

%------------------------------------------------------------------------------
\subsection{Mock dSphs}
\label{subsec:mock}

\begin{table}
\begin{center}
\begin{tabular}{ccccc}
%\begin{tabular}{p{1cm}p{1cm}p{1cm}p{1cm}p{1cm}}
dSph        & $d$ [kpc] & $r_{e}$ [pc] & $(l,\,b)$&  $v_{\text{dSph}}$ [km/s] \\ \hline
Draco       &      76 &          221 & (86.4,\,34.7)&  -292 \\
Ursa\,Minor  &      76 &          181 & (105,\,44.8) &  -247  \\ \hline
  \end{tabular}
\caption{{\sl
The properties of the galaxies which are given in \citet{McConnachie:2012vd} and references therein.
The distances to the dSph are denoted by $d$ and 
$(l,\,b)$ are the galactic longitude and latitude of each dSph. 
The bulk line-of-sight velocity of each dSph is given by $v_{\text{dSph}}$\,\citep{Falco:1999vs, Young:1999kh}.
The negative sign represents that the dSph is moving toward the sun.
}}
\label{tb:dSphProperty}
\end{center}
\end{table}

We construct the mock stellar data based on the classical dSphs data (where the number of the observed member stars is $300\--500$) with a large $J$-factor.\footnote{
We also stress the importance of the ultra-faint dSphs which often have larger $J$-factors than classical dSphs.
The current limited stellar data does not provide a solid value of the $J$-factor
and therefore the future spectrographic survey and our analysis for the foreground stars should play essential roles.
However, other systematic errors may also largely contribute to the dark matter halo estimation and 
in this paper, in order to emphasize the effect of the foreground contamination,
we only consider the classical case.
The analysis of the ultra-faint dSphs will be investigated in our forthcoming paper.
}
This amount of the stellar data provides relatively well-determined dark matter profile
and therefore, they are especially important when one considers the conservative sensitivity lines.
Here we consider Draco and Ursa\,Minor dSphs since they have the largest $J$-factors among the classical dSphs and primary targets of the future spectrographs.
Here, we extract the halo data from Draco and Ursa\,Minor observations and 
use it for the input of the mock generator (see Table~\ref{tb:dSphProperty} for the properties of each dSph).

The mock stellar data of the dSphs is constructed  by two steps.
At the first step, we generate the member stars with the colour and chemical information by using a stellar evolution model deduced from the current photometric and spectroscopic data.
The next step is to assign to each star the information about the position and velocity consistent with the phase-space distribution function determined by the input dark matter potential.

\begin{table*}
\begin{center}
  \begin{tabular}{cccccccc}
dSph
& $\log_{10}(\rho_{s}/[M_{\bigodot}/\text{pc}^3]) $
& $\log_{10}(r_s/[\text{pc}])$
& $\alpha$ & $\beta$ & $\gamma$ & $\log_{10} (1-\beta_{\text{ani}})$ & \\ \hline
Draco\,1      & -2.05  & 3.96 & 2.78 & 7.78 & 0.675 &  0.130 &\\
Draco\,2      & -1.52  & 3.15 & 2.77 & 3.18 & 0.783 & -0.005 &\\
Ursa\,Minor  & -0.497 & 2.60 & 1.64 & 5.29 & 0.777 & -0.475 &\\
\hline
 \end{tabular}
\caption{{\sl
The dark matter halo parameter of each dSph.
}}
\label{tb:dSphParameter}
\end{center}
\end{table*}

In the first step, the synthetic colour-magnitude diagrams for the Draco and Ursa\,Minor are generated by utilizing the latest version of the PERSEC isochrones\,\citep{bressan12}. 
We first randomly draw initial masses of stars from the Salpeter initial-mass function\,\citep{Salpeter:1955it}. 
Ages and metallicities are also randomly drawn from assumed distributions.
The ages of the stars are assumed to be randomly distributed in the range 12.6-13.2 Gyrs for both galaxies.
The stellar metallicities ([Fe/H]) are assumed to follow a Gaussian distribution with a mean and dispersion -1.9 (-2.1) and 0.5 (0.5) dex, respectively, for Draco (Ursa\,Minor), which approximately reproduce the observed metallicity distributions in these galaxies\,\citep{kirby11}.
The present-day absolute magnitude, temperature, and surface gravity are then assigned to each star based on the PERSEC isochrones for the given initial mass, age, and metallicity.
The apparent magnitudes are obtained by taking into account the distance modulus of 19.40, which corresponds to the distance of 76 kpc, for both Draco and Ursa\,Minor\,\citep{mcconnachie12}.
Finally, the photometric errors, which are assumed to increase toward fainter magnitudes with a cubic polynomial, are assigned to the apparent magnitudes. 
The number of stars in each galaxy is adjusted to yield the total luminosity approximately consistent with the observed luminosity.

After constructing the member star mock, we next randomly assign the position and velocity of each star by using the kinematical distribution of the dSph (and finally add the bulk velocity $v_{\text{dSph}}$).
The stellar distribution consistent with the dark matter potential is obtained  by the method in \citet{1991MNRAS.253..414C} in which we assume the constant velocity anisotropy parameter. 
\footnote{
In order to avoid to make unphysical distribution, we have slightly modified the input stellar distribution form by introducing the small parameter $\gamma_{*}$: 
$\nu_{*}(r) \rightarrow (3-\gamma_{*})/4 \pi r_e^3 (r/r_{e})^{-\gamma_{*}} (1+r^2/r_{e}^2)^{-(5-\gamma_{*})/2}$.
In the analysis, we have set $\gamma{*} = 0.1$ and confirmed that the fluctuations of the reproduced number distribution and dispersion curves are negligible.
}
As the input dark matter halo, we adopt two types of the halo profile considering `Draco-like' and  `Ursa\,Minor-like' dSph.
For Draco-like dSphs, we estimate the halo profile by utilizing 
the current kinematical stellar data given by MMT/Hectochelle observations\,\citep*{2015MNRAS.448.2717W}.
We obtain the best fit parameters of the halo by the same method in \citet{Geringer-Sameth:2014yza}
as shown in the first line in Table~\ref{tb:dSphParameter} (`Draco\,1').\footnote{
The fit uses the $\sim450$ kinematical data. 
}
However, the best fit data for Draco usually gives large $r_s$ ($\sim 10\,\text{kpc}$).
In fact, the amount of the foreground contamination is not obvious for the current observational data.
As one can see later, foreground contamination gives a large velocity dispersion at outer region 
and leads overestimation of the halo size.
Therefore, we also adopt another fit parameter with smaller $r_s$ and a good chi-square 
($0.1\,\%$ larger than the best fit chi-square) 
as shown in the second line of Table~\ref{tb:dSphParameter} (`Draco\,2').
For Ursa\,Minor-like dSphs, on the other hand,
because the kinematical data of the Ursa\,Minor is not available,
we adopt the median values of Ref\,\citep{Geringer-Sameth:2014yza} 
as the input parameter of the halo which is shown in the third line of Table~\ref{tb:dSphParameter} (`Ursa\,Minor').

The non-member stars belonging to the Milky Way galaxy are also included in the mock data. 
These foreground stars are generated from the Besan\c{c}on model\,\citep{Robin:2004qd}.
The generator provides the stellar population of the Milky Way galaxy including the thin disc, thick disc, bulge and halo component with its velocity, age, luminosity, colour, chemical components, effective temperature and surface gravity.
Using the Besan\c{c}on model, 
we generate the foreground stars with spatially uniform distribution in the region of interest.

%------------------------------------------------------------------------------
\subsection{Spectrograph}

\begin{table*}
\begin{center}
  \begin{tabular}{cccccc}
    $\theta_{\text{ROI}}$ [degree] & $i_{\text{max}}$ [mag] & $dv$ [km/s]& $d$[Fe/H] & $d\log_{10}(g/[\text{cm}/\text{s}^2])$ & $ dT_{\text{eff}}$ [K] \\ \hline
    $0.65,\,1.3$ & $19.5,\,21,\,21.5$       &    $3.0$      & $ 0.5$    & $0.5$                                  & $500$                  \\ \hline
  \end{tabular}
\caption{\sl
The capability of the spectrograph.
$\theta_{\text{ROI}}$ is the radius of the region of interest.
}
\label{tb:Spectrograph}
\end{center}
\end{table*}

As for the future spectrograph, we consider the Prime Focus Spectrograph (PFS)
mounted on the 8.2m Subaru telescope.
PFS is the next generation spectrograph of the SuMiRe project\,\citep{2014PASJ...66R...1T, 2015JATIS...1c5001S,Tamura:2016wsg},
which has a large field-of-view ($\sim 0.65$ degrees) and 2394 fibres.
The project has a survey plan for the classical dSphs 
(Fornax, Sculptor, Draco, Ursa\,Minor, and Sextans) and 
aims to start the science operation from $2019 \-- 2020$.
The spectrograph has the three-colour arms which cover blue, 
red and near infrared wavelengths with the resolution $\lambda/\delta \lambda$ of 2500, 3200, and 4500, respectively. 
Moreover, the spectrograph has the medium-resolution option for the red-arm ($\lambda/\delta \lambda = 5000$),
where the velocity precision $dv$ of $3 \unit{km/s}$ is expected.\footnote{
The medium resolution mode ($\lambda =7100\--8850 \unit{\AA}$) covers the Calcium triplet and $\alpha$ element lines.
}
In addition, utilizing the absorption line spectrum in the wide wavelength coverage, 
it is expected that the detailed stellar parameters (e.g. effective temperature, surface gravity, metallicity) can be obtained with high accuracies.
In the survey plan, the star with $i > 21$ will be observed with an integration time of a few nights,
while the current observation measures stars above $i \sim 19.5 $ spending several nights to cover the dSph region.
%and therefore much larger stellar will be observed by PFS.

We note that other future spectrographs planned to start operation 
in the next several years, such as 
MOONS at VLT\,\citep{Cirasuolo:2012tw},
WEAVE at 4.2 m William Herschel Telescope\,\citep{Balcells:2010ck}, 
DESI at 4 m Mayall telescope\,\citep{Levi:2013gra},
or 4MOST at VISTA telescope\,\citep{deJong:2012nj}, could also provide 
velocity measurements with similar quality ($ dv  \leq 3 \unit{km/s}$). 
Compared to these instruments, the Subaru/PFS have both a wide
field-of-view and the large aperture telescope 
and thus is particularly suitable for measuring velocities 
in the outer region of the classical dSphs down to a fainter
magnitude ($i<21.0$). Nevertheless, the analyses presented below 
provide a benchmark for improvement/limitation of stellar kinematic
data taken with these instruments as well as with a next-generation 
instrument such as the Maunakea Spectroscopic Explore
(MSE;\ \citet{2016SPIE.9906E..2JS}).

For the region of interest,
we assume one-pointing observation (a radius of 0.65 degrees) and four pointing observation (1.3 degrees). 
We also assume that the spectrograph can measure
the recession velocity $v$, metallicity [Fe/H], effective temperature $T_{\text{eff}}$, 
and surface gravity $g$ for each star with the accuracies given in Table~\ref{tb:Spectrograph}, which are used to eliminate the foreground stars.\footnote{
We check the effect of the velocity resolution and find it small because the dispersion is given by $\sqrt{\sigma^{2}_{r} + dv^2}$. For classical dSph case, $\sigma_{r} \gtrsim 10$ (km/s) and the effect of $dv$ negligibly contributes to the $J$-factor uncertainty.
}
To simulate the detector resolution,
the mock data is smeared by the normal distribution functions
with the respective resolution widths given in Table~\ref{tb:Spectrograph}.
As the depth of the survey depends on the exposure time, 
we adopt three cases of the upper bound of 
the magnitude ($i_{\text{max}} = 19.5,\,21,\,21.5$).
In the first case,
we demonstrate  the current sensitivity reach
and the effect of the foreground contamination.
The maximum magnitude $i_{\text{max}} = 19.5$ is chosen
to reproduce the current number of stellar data ($\sim 300$).
The second case ($i_{\text{max}}=21$) corresponds to the expected reach of PFS one-pointing observation
with an integration time of a few nights.
As for the multi-pointing case, we consider the option with a wider region of interest ($\theta_{\text{ROI}} = 1.3 \unit{degree}$).
The final case ($i_{\text{max}} = 21.5$) is for a deeper survey with an integration time of several nights,
in which we consider only the one-pointing case.

\begin{table*}
\begin{center}
  \begin{tabular}{ccccccc}
  dSph
& $\theta_{\text{ROI}}$ [degree] 
& $i_{\text{max}}$ [mag] 
& $v_{\text{lower}}$ [km/s] 
& $v_{\text{upper}}$ [km/s] 
& $N_{\text{Mem}}$
& $N_{\text{FG}}$
\\ \hline
Draco 1 \& 2 & $0.65$ & $19.5$  & $-350$ & $-230$ &  260 &  16 \\
             &        & $21$    &        &        &  900 &  37 \\
             &        & $21.5$  &        &        & 1140 &  43 \\
             & $1.3$  & $21$    &        &        &  940 & 150 \\ \hline
Ursa\,Minor  & $0.65$ & $19.5$  & $-310$ & $-190$ &   290 &  10 \\ 
             &        & $21$    &        &        &  1100 &  33 \\ 
             &        & $21.5$  &        &        &  1400 &  41 \\ 
             & $1.3$  & $21$    &        &        &  1130 & 140 \\ \hline
  \end{tabular}
\caption{\sl
The status of the mock dSphs.
The averaged number of the member (foreground) stars after the cuts are given by $N_{\text{Mem}}$ ($N_{\text{FG}}$).
See the text for the details of the cuts.
}
\label{tb:Cut}
\end{center}
\end{table*}

%------------------------------------------------------------------------------
\subsection{Cut}
\label{subsec:cut}

\begin{figure}
\begin{center}
\includegraphics[scale=0.37]{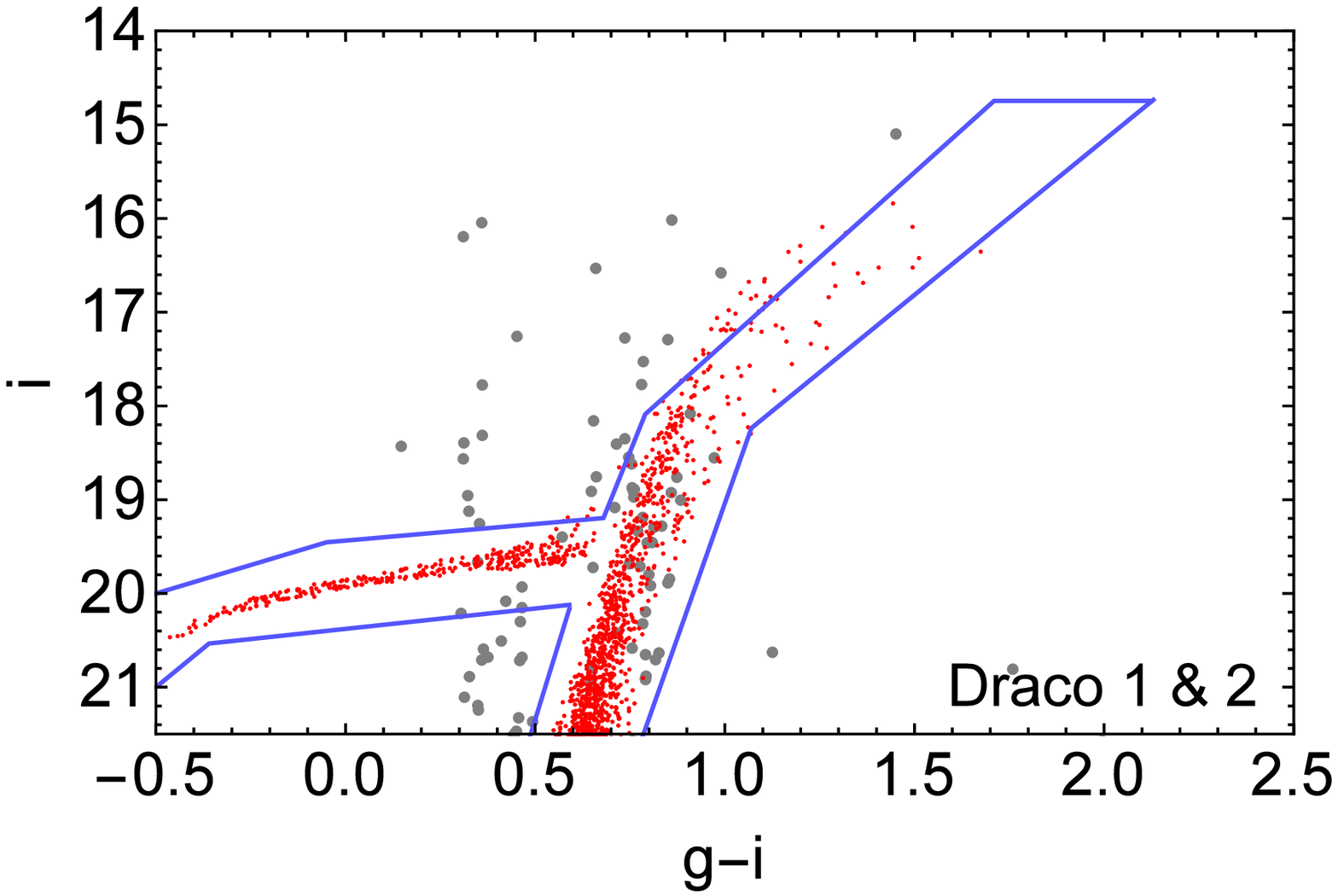}
~~
\includegraphics[scale=0.37]{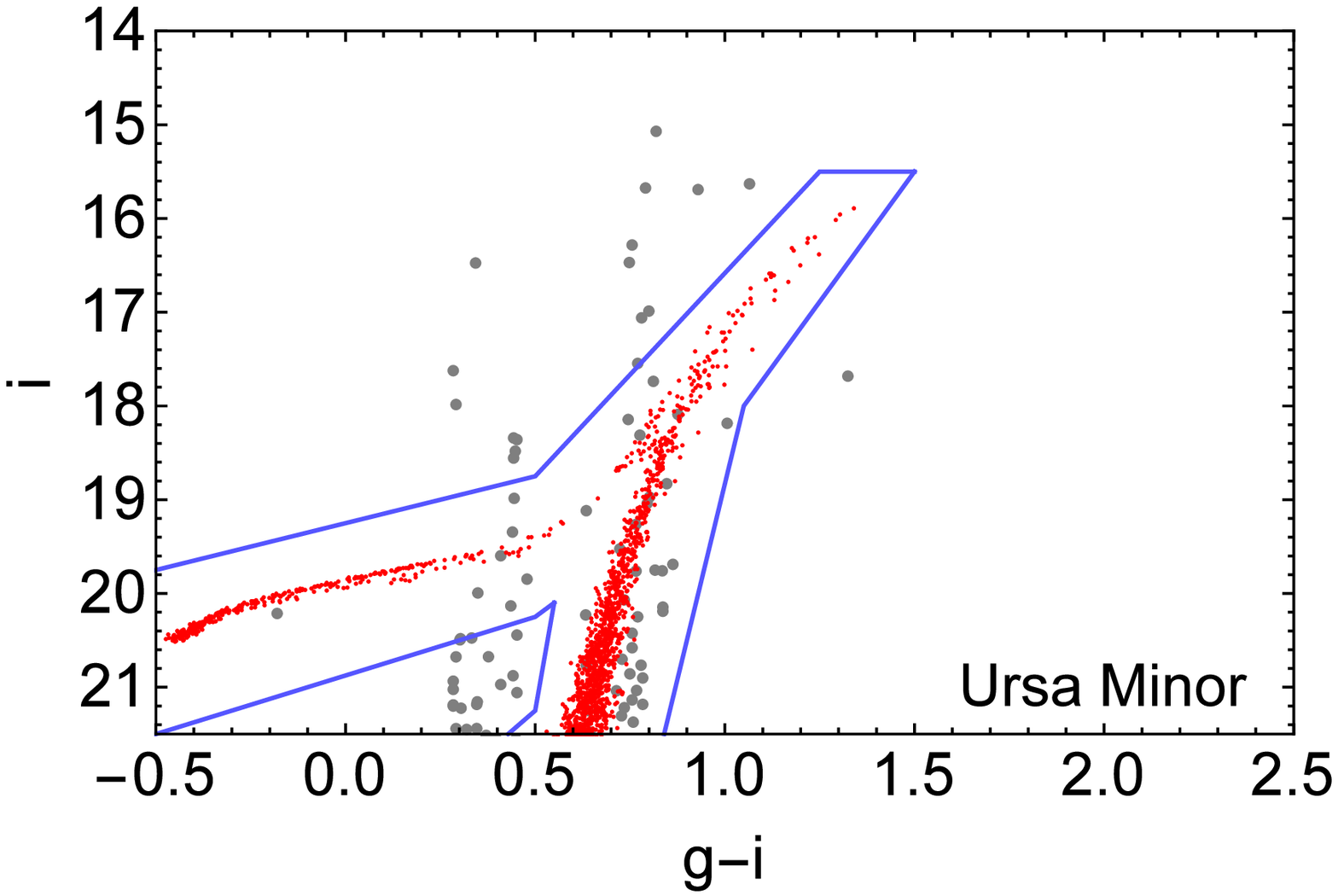}
\caption{\small\sl 
The colour-magnitude map for Draco (top) and Ursa\,Minor (bottom).
We impose the colour-magnitude cut by the blue lines.
The red (grey) dots show the members (foreground) stars.
The stars on the map are residuals after the cuts of the ROI, velocity, and $\log g$.
For Draco, we draw the colour cut referring \citet{2015MNRAS.448.2717W}.
}
\label{fig:dSphCut}
\end{center}
\end{figure}

The mock data obtained in Sec.\,\ref{subsec:mock} contain a large number of the foreground stars.
To reduce the foreground contamination, we impose cuts on the raw mock data.\footnote{
Standard analysis of $J$-factor estimation
uses stars selected  according to their membership probabilities, 
which can be a more sophisticated selection than the cut method introduced in this section.
The distribution of the selected data can be different from that obtained by the cut.
Therefore it is worth comparing the results with those given by the standard analysis.
We discuss it in appendix B.
}
First, the spatial 
($r < d \sin \theta_{\text{ROI}}$) cut should be imposed 
because of the limited region of interest. 
Here $d$ denotes the distance of each dSph and $\theta_{\text{ROI}}$ is the angular radius of the region of interest.
We further optimize the foreground contamination by imposing following cuts:
\begin{itemize}
\item $v_{\text{lower}} < v < v_{\text{upper}}$\ ,
\item $0.2 < \log_{10} (g/[\text{cm}/\text{s}^2]) < 3.7$\ ,
\end{itemize}
as well as the cut on the colour-magnitude diagram as shown in Fig.\,\ref{fig:dSphCut}.
Here, $v_{\text{lower}},\,v_{\text{upper}}$ are given in Table~\ref{tb:Cut} for each dSph.
We choose the velocity bound to include most of stars in clumps (by eyes).
A harder cut can be imposed to reduce the fraction of the foreground stars.
However, in that case, since the scattered member stars are also eliminated,
the reconstructed velocity distribution may be biased.
The effect of this hard cut is beyond the scope of this paper and we do not go into further details.
The number of the member and foreground stars after imposing the cuts is also shown in the table.
Since the bulk velocity of the dSphs is largely different from the Milky Way stellar halo component (see Fig\,.\ref{fig:FGdist} in appendix A), the velocity cut works most efficiently.
We also note that because the surface gravity reflects the absolute magnitude, 
the surface gravity cut can efficiently eliminate the faint foreground stars from the member stars with brighter absolute magnitude.
As most of the residual foreground stars belong to the Milky Way stellar halo component, additional cuts using the metallicity or effective temperature are not efficient
because the member stars are indistinguishable from the halo stars in terms of these quantities for the accuracy given in Table~\ref{tb:Spectrograph}.
Rather, these additional cuts eliminate the member stars scattered by the detector resolution by 
$15\,\%$ level.

\subsection{Velocity dispersion}
The velocity dispersion along the line-of-sight can be obtained from Eq.(\ref{eq:Jeans1})and (\ref{eq:dispersion}).
Here we note that Eq.(\ref{eq:Jeans1}) has 
a general solution\,\citep{1994MNRAS.270..271V,2005MNRAS.363..705M}.
When the $\beta_{\text{ani}}$ is constant, the radial velocity dispersion can be expressed as
\begin{eqnarray}
\sigma^{2}_{r}(r) = 
\frac{1}{\nu_{*} (r) }
\int^{\infty}_{r} 
\nu_{*} (r')
\left(\frac{r'}{r}\right)^{2 \beta_{\text{ani}}}
\frac{G M(r')}{r'^2} 
dr'\ .
\label{eq:dispgen}
\end{eqnarray}
Combining Eq.(\ref{eq:dispersion})-(\ref{eq:dispgen}),
we convert $\sigma^{2}_{r}(r)$ to the projected velocity dispersion $\sigma^{2}_{l.o.s}(R)$,
which is calculated by inputting
halo parameters $\alpha,\,\beta,\,\gamma,\,\rho_s,\,r_s$ and velocity anisotropy $\bani$.
We also note that $M(r) \equiv \int^{r}_{0} 4 \pi r'^2 \rho (r') dr'$ under the spherical assumption 
and we define $\Sigma_{*}(R) = (1/\pi r_{e}^{2})\,(1 + (R/r_{e})^{2})^{-2}$ for the Plummer profile.

%------------------------------------------------------------------------------
\subsection{Likelihood}
\label{subsec:pfslikelihood}
The likelihood function is constructed taking the foreground contamination into account.
To avoid the binning dependence, we perform the unbinned analysis
by setting the likelihood as follows:
\begin{eqnarray}
-2 \ln {\cal L} = -2 \sum_{i} \ln ( s f_{\text{Mem}} (v_{i}, R_{i}) + (1-s) f_{\text{FG}} (v_{i}, R_{i}) )\ ,
\label{eq:likeli}
\end{eqnarray}
where
$s$ is the membership fraction parameter and
$f_{\text{Mem}}(v,R)$ ($f_{\text{FG}}(v,R)$) is the distribution function of the member (foreground) stars.
The index $i$ runs all the stars in the mock data set.
The distribution functions are defined by
\begin{eqnarray}
\hspace{10pt} f_{\text{Mem}}(v,R) = 2 \pi R \Sigma_{*}(R)\,C_{\text{Mem}}\,{\cal G}[v;\,v_{\text{Mem}},\,\sigma_{l.o.s} (R)]\ ,\label{eq:fmem}\\
f_{\text{FG}}(v,R) = 2 \pi R\,C_{\text{FG}}  {\cal G}[v;\,v_{\text{FG}},\,\sigma_{\text{FG}}]\ ,\hspace{55pt}
\end{eqnarray}
where $\sigma_{\text{FG}}$ does not depend on $R$.
Here, we assume that both of the velocity distributions can be approximated by a single Gaussian,
and hence, ${\cal G}[x;\,\mu,\sigma]$ denotes the Gaussian distribution of a variable $x$ 
with a mean value, $\mu$ and a standard deviation, $\sigma$.
The parameter $v_{\text{Mem}}$ represents the bulk velocity of the dSph
while $v_{\text{FG}}$ is (dominantly) controlled by the bulk velocity of the foreground halo component,
which are treated as nuisance parameters in the following analysis.
$C_{\text{Mem}}$ and $C_{\text{FG}}$ are the normalization correction factor, under which the distribution functions satisfy
\begin{eqnarray}
\int^{r_{\text{ROI}}}_{0}dR \int^{v_{\text{upper}}}_{v_{\text{lower}}} dv f_{\text{Mem}}(v,R) = 1\ , \\
\int^{r_{\text{ROI}}}_{0}dR \int^{v_{\text{upper}}}_{v_{\text{lower}}} dv f_{\text{FG}}(v,R) = 1\ ,
\end{eqnarray}
where $r_{\text{ROI}} \equiv d\,\sin \theta_{\text{ROI}}$.
We note that the free parameter $v_{\text{Mem}}$ always converges to the input bulk velocity $v_{\text{dSph}}$.

Before the fit, 
the information about $v_{\text{FG}},\,\sigma_{\text{FG}}$ can be extracted by utilizing the data set in the control region, i.e.,
the data set with $ v < v_{\text{lower}}$ or $v > v_{\text{upper}}$.\footnote{
The control region can also be taken by the spatial position, setting an annulus centred at the dSph galaxy.
However, we have found that the fraction of the member star in the annulus is not negligible and 
therefore we decided to use $v$ to define the control region.
}
Performing a fit to the control region, 
the best fit value and standard deviation of $v_{\text{FG}},\,\sigma_{\text{FG}}$ 
($v_{\text{FG0}},\,\sigma_{\text{FG0}}$, $dv_{\text{FG}},\,d\sigma_{\text{FG}}$)
can be obtained. 
We use this information as a prior for $v_{\text{FG}},\,\sigma_{\text{FG}}$ 
by multiplying 
${\cal G}[v_{\text{FG}};\,v_{\text{FG0}},\,dv_{\text{FG}}]\,{\cal G}[\sigma_{\text{FG}};\,\sigma_{\text{FG0}},\,d\sigma_{\text{FG}}]$
to the likelihood function ${\cal L}$ in Eq.(\ref{eq:likeli}).
Here we emphasize that to construct the foreground distribution function, the amount of the foreground data is essential.
For instance, the number of the stars after the colour and ROI cuts is $\sim 2000$ for the one-pointing case.
Such a large number of stars can only be accessed by the spectrograph with many fibres, like PFS.
The detailed method to estimate the foreground distribution function is given in appendix A.
\footnote{
We here note that this fit procedure is essentially the same as the one 
proposed in the appendix of \citet{Bonnivard:2015vua}
as an alternative procedure for their main analysis. 
However, the main difference is that we derive the foreground distribution function by utilizing the control region, which is introduced as the prior function in the likelihood. 
As another difference, we use the information about the metallicity only for the cut and do not include them in the likelihood. 
This is because the metallicities of the member stars and foreground halo components are highly degenerate with each other.
%Moreover, we use the information about metallicity only for the cut and do not include them in the likelihood. This is because the metallicity of the member stars and foreground halo components are highly degenerate and does not work efficiently in the likelihood.
}
We check that 
the parameters of the foreground distribution ($v_{\text{FG}}$, $\sigma_{\text{FG}}$) converges within $\sim 1$ sigma width of the prior on average.

%The likelihood (multiplied by the foreground priors) is maximized under 
The Bayesian posterior probability function of 
the five free parameters of the dark matter halo
($\rho_s,\,r_s,\,\alpha,\,\beta,\,\gamma$), 
one velocity anisotropy parameter $\beta_{\text{ani}}$
and 
four nuisance parameters in the likelihood function ($s$, $v_{\text{Mem}},\,v_{\text{FG}},\,\sigma_{\text{FG}}$)
is obtained from the likelihood function in Eq.(\ref{eq:likeli}) multiplied by the foreground priors.
We perform the Metropolis-Hastings (MH) algorithm\,\citep{Metropolis:1953am,Hastings:1970aa} of the Markov Chain Monte Carlo (MCMC) method.
Once one properly tunes the MCMC process
(such as the number of the burn-in step, the sampling step, and length of the chain),
the sampling set of the MCMC reflects the probability density of the likelihood function.
Accumulating ${\cal O}(10^5)$ samples for each data set,
we search the halo parameters 
under the flat/log-flat priors within the range of
$-4 < \log_{10} (\rho_{s}/[\text{M}_{\bigodot}/\text{pc}^3]) < 4 $, 
$ 0 < \log_{10} (r_{s}/[\text{kpc}]) < 5 $,
$0.5 < \alpha < 3  $,
$ 3 < \beta < 10  $,
$ 0 < \gamma < 1.2  $ and 
$-1 < \log_{10} (1-\beta_{\text{ani}}) < 1 $,
which are the same criteria of\,\citet{Geringer-Sameth:2014yza}.

%------------------------------------------------------------------------------
%-----------------------          Results          ----------------------------
%------------------------------------------------------------------------------
\section{Results}
\label{sec:result}

\renewcommand{\arraystretch}{1.3}
\begin{table*}
\begin{center}
\scalebox{0.85}[0.85]{
  \begin{tabular}{ccc|ccc}
\multicolumn{3}{c|}{Condition} & \multicolumn{3}{|c}{ $\log_{10} (J/[\text{GeV}^2/\text{cm}^5])$ } \\
\hline 
  Mock
& $\theta_{\text{ROI}}$ %[degree] 
& $i_{\text{max}}$ %[mag] 
& Input
& Contaminated
& Our fit
\\ 
\hline
%Draco\,1          &     $0.65$     &     $19.5$        &     $18.94$     &     $ 19.41 \ ^{+0.14}_{-0.14}\ \pm 0.10$     &     $ 18.72 \ ^{+0.16}_{-0.15}\ \pm 0.13$          \\
Draco\,1 & $0.65$ & $19.5$ & $18.94$ & $19.40 ^{+0.13}_{-0.13}\ \pm 0.15$ & $18.74 ^{+0.17}_{-0.16}\ \pm 0.16$ \\
 & $ $ & $21$ &  & $19.41 ^{+0.08}_{-0.08}\ \pm 0.09$ & $18.71 ^{+0.11}_{-0.10}\ \pm 0.08$ \\
 & $ $ & $21.5$ &  & $19.38 ^{+0.07}_{-0.08}\ \pm 0.10$ & $18.71 ^{+0.10}_{-0.09}\ \pm 0.09$ \\
 & $1.3$ & $21$ &  & $19.42 ^{+0.06}_{-0.05}\ \pm 0.06$ & $18.75 ^{+0.10}_{-0.09}\ \pm 0.08$ \\
\hline                        
%Draco\,2          &     $0.65$     &     $19.5$        &     $18.88$     &     $ 19.47 \ ^{+0.16}_{-0.14}\ \pm 0.09$     &     $ 18.98 \ ^{+0.18}_{-0.16}\ \pm 0.15$          \\
Draco\,2 & $0.65$ & $19.5$ & $18.88$ & $19.48 ^{+0.14}_{-0.13}\ \pm 0.17$ & $18.87 ^{+0.17}_{-0.15}\ \pm 0.12$ \\
 & $ $ & $21$ &  & $19.39 ^{+0.09}_{-0.09}\ \pm 0.10$ & $18.84 ^{+0.11}_{-0.09}\ \pm 0.07$ \\
 & $ $ & $21.5$ &  & $19.38 ^{+0.08}_{-0.08}\ \pm 0.08$ & $18.83 ^{+0.09}_{-0.08}\ \pm 0.08$ \\
 & $1.3$ & $21$ &  & $19.43 ^{+0.06}_{-0.05}\ \pm 0.06$ & $18.84 ^{+0.09}_{-0.08}\ \pm 0.06$ \\
\hline
%Ursa\,Minor      &     $0.65$     &     $19.5$        &     $19.03$     &     $ 19.33 \ ^{+0.13}_{-0.11}\ \pm 0.09$     &     $ 19.14 \ ^{+0.15}_{-0.12}\ \pm 0.09$           \\
Ursa\,Minor & $0.65$ & $19.5$ & $19.03$ & $19.38 ^{+0.13}_{-0.11}\ \pm 0.12$ & $19.12 ^{+0.15}_{-0.12}\ \pm 0.09$ \\
 & $ $ & $21$ &  & $19.30 ^{+0.06}_{-0.06}\ \pm 0.07$ & $19.11 ^{+0.12}_{-0.08}\ \pm 0.05$ \\
 & $ $ & $21.5$ &  & $19.28 ^{+0.06}_{-0.05}\ \pm 0.05$ & $19.09 ^{+0.10}_{-0.07}\ \pm 0.05$ \\
 & $1.3$ & $21$ &  & $19.45 ^{+0.05}_{-0.05}\ \pm 0.03$ & $19.10 ^{+0.13}_{-0.08}\ \pm 0.06$ \\
\hline                        
 \end{tabular}
}
\caption{\sl
The resultant $J$-factors 
calculated within an angular radius of $0.5^{\circ}$.
We produce 50 mocks and we give the mean (the first values) and averages of the error bars (the first uncertainties). The standard deviation of the median values is also put on the second uncertainty. The $J$-factors calculated by the input parameters are given in the `{\sl Input}' column. The `{\sl Contaminated}' column shows the results where all the data after the cut are considered as the member. The `{\sl Our fit}' column shows the fit results obtained by using the likelihood of Eq.(\ref{eq:likeli}). 
}
\label{tb:result}
\end{center}
\end{table*}
\begin{figure}
 \begin{center}
  \includegraphics[width=80mm,angle=0]{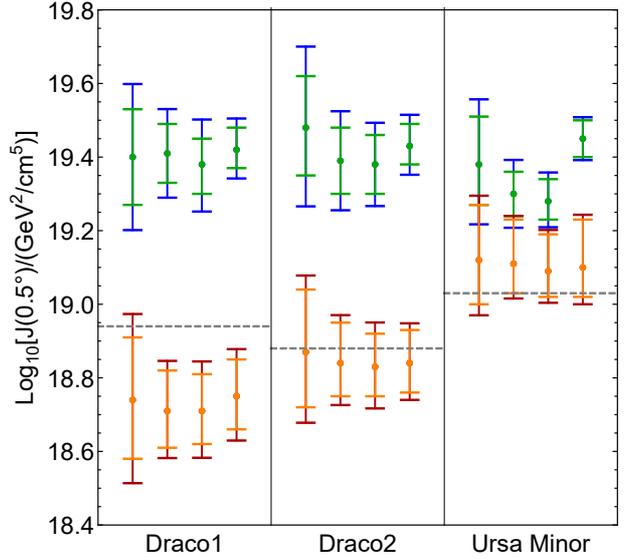}
\caption{\small\sl
The $J$-factors in Table~\ref{tb:result} are plotted. The orange dots show the $J$-factor estimation of `{\sl Our fit}', while the green dots represent the `{\sl Contaminated}' results. The orange and green bars are the first uncertainties in Table~\ref{tb:result}. The red and blue bars are the errors where the standard deviation of the median values (the second uncertainties in Table~\ref{tb:result}) is added by square root sum. The grey dashed lines show the input values. For each dSph, the first three bars from the left correspond to the case of $i_{\text{max}} = 19.5,\,21,\,21.5$ with $\theta_{\text{ROI}} = 0.65$ and the last bar is $i_{\text{max}} = 21$ with $\theta_{\text{ROI}} = 1.3$ in Table~\ref{tb:result}
}
 \label{fig:Jfig}
 \end{center}
\end{figure}

\begin{figure}
 \begin{center}
  \includegraphics[width=70mm,angle=0]{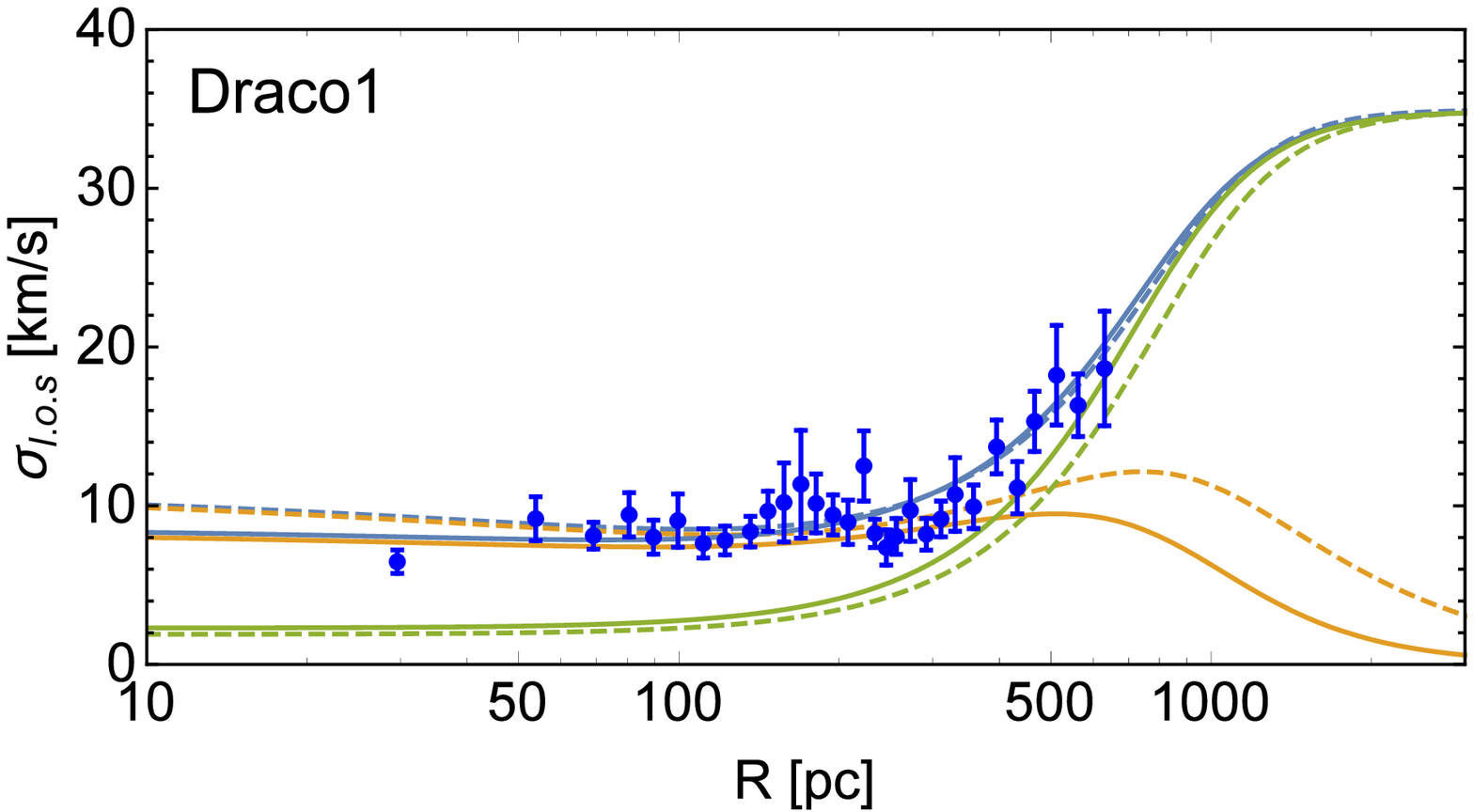}
\\
  \includegraphics[width=70mm,angle=0]{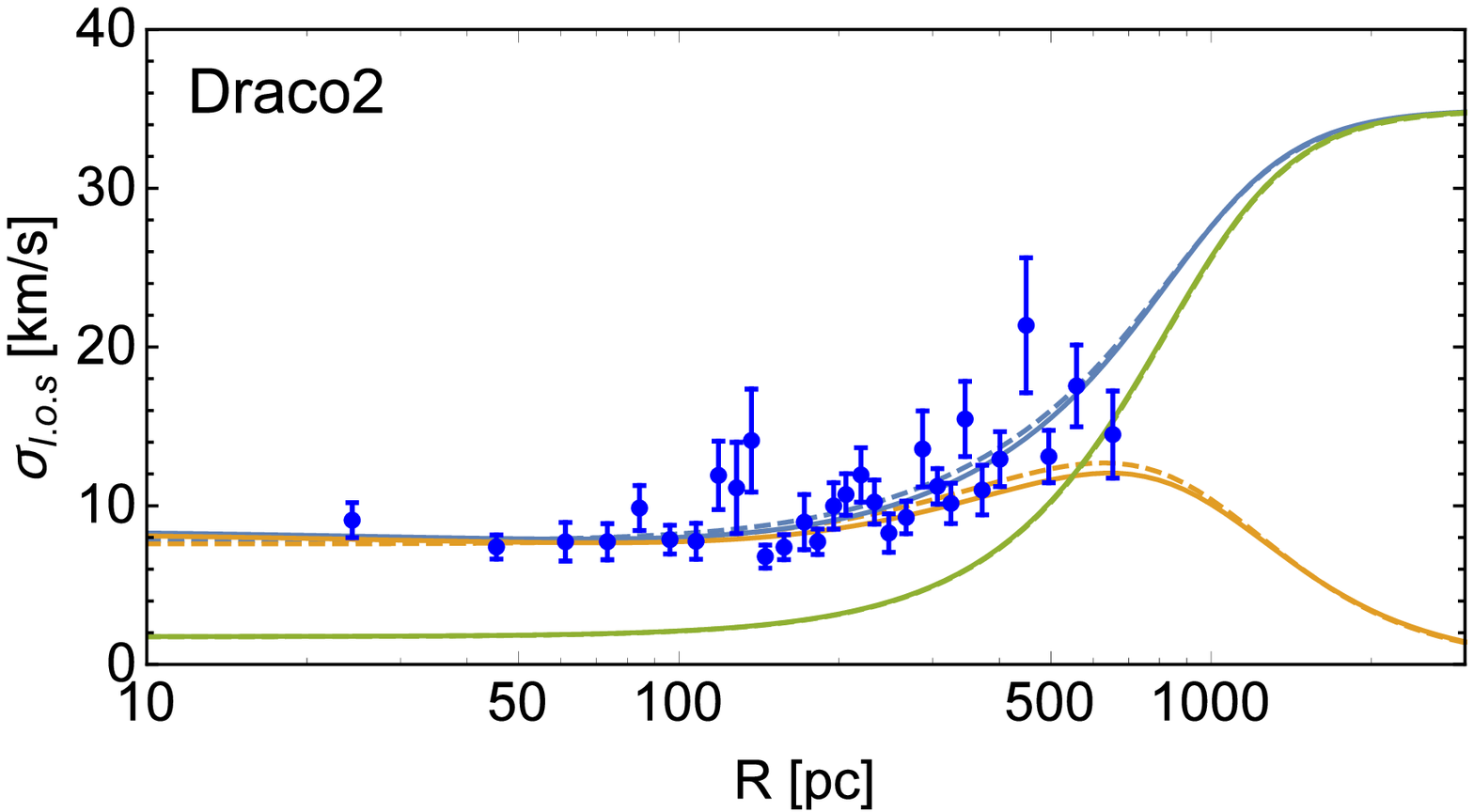}
\\
  \includegraphics[width=70mm,angle=0]{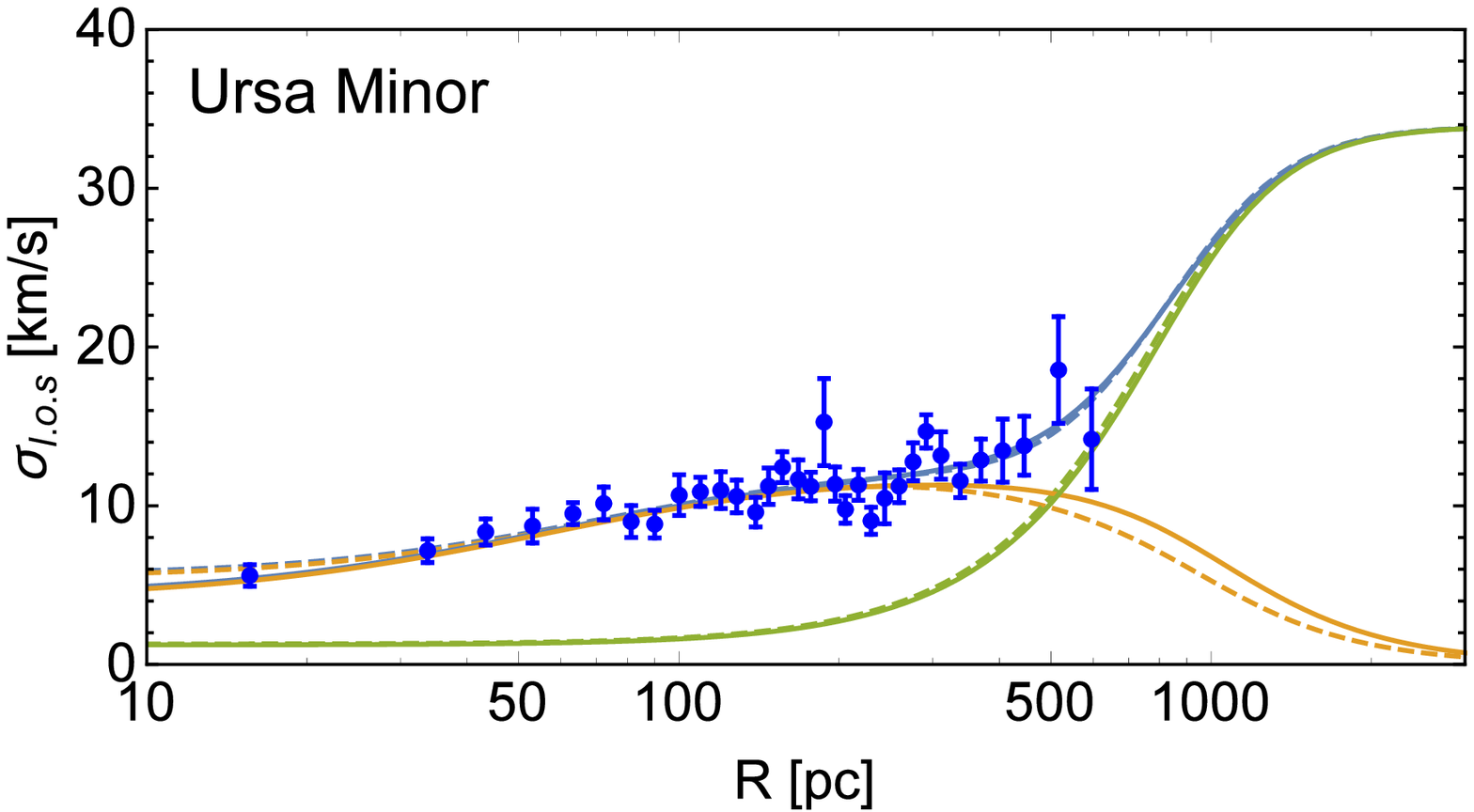}
 \end{center}
 \caption{\small\sl
The dispersion curve of $\theta_{\text{ROI}} = 0.65,\,i_{\text{max}} = 21$ case of 
Draco\,1 (top), Draco\,2 (middle), and Ursa\,Minor (bottom).
The binned dispersions of the mock data are shown by the blue dots with error bars.
The solid blue line shows the dispersion curve obtained from the best fit parameter.
The solid orange (green) line shows the member (foreground) contribution to the dispersion curve.
The dashed orange line is obtained from the input parameter of the dSph dark matter halo.
The dashed green line is the curve obtained from the mean value of the foreground prior,
while the dashed blue line shows the sum of them.
}
 \label{fig:disp}
\end{figure}

Table~\ref{tb:result} and Fig.\,\ref{fig:Jfig} show the results of the fit.
We produce 50 mocks for each case and average the median values of $\log_{10} (J/[\text{GeV}^2/\text{cm}^5])$.
The error bar of the $J$-factor for each mock sample is estimated by comparing the median and 68\,\% quantile.
In the table, we give the averages of the error bars at the first uncertainties.
The standard deviation of the median values is also put on the second uncertainty, 
which reflects the statistical deviation of the sample quality.
All $J$-factors are calculated in $\Delta \Omega = 2.4 \times 10^{-4}\,\text{sr}$, corresponding to the angular radius of $0.5 \,\text{degree}$, which is the standard size for the $J$-factor calculation\,\citep{Ackermann:2015zua}.\footnote{
We also check how the error of the $J$-factor are affected by changing this integration angle.
For Draco\,1, Draco\,2, and Ursa\,Minor case,
we find that the integration angle of $\sim 0.3$ degree gives the smallest error bar for the 
$J$-factor estimations, which corresponds to a scale of  $\sim 2 r_{e}$.
A wider integration angle gives a larger error and 
the error bar of  0.5 degree is larger than that minimum by a factor of two at most.
Therefore, more robust estimations can be derived for these dSphs by tuning the integration angles.
}
To eliminate the fluctuation due to the halo truncation, 
as we have mentioned in Sec.\,\ref{subsec:systematics},
we fix the size of the halo truncation at $2000\,\text{pc}$.\footnote{
We give the effect by changing this truncation radius in appendix C.
}
The $J$-factors calculated by the input parameters are given in the `{\sl Input}' column. The `{\sl Contaminated}' column shows the results where all the data after the cut are considered as the member (meaning that we fix $s = 1$ in Eq.(\ref{eq:likeli})). The `{\sl Our fit}' column shows the results obtained by using the likelihood of Eq.(\ref{eq:likeli}).

As one can see from the figure, the results obtained from our likelihood successfully reproduces the input value. 
%For the $i_{\text{max}} = 19.5$ case, 
%since the observed kinematical data are rather limited, not only the estimation error (the first uncertainty) 
% but also the sampling fluctuation (the second uncertainty) are larger than the $i_{\text{max}} \geq 21$ case.
For $i_{\text{max}} = 21$ case, the estimation error (the first uncertainty) is about $20 \-- 35$ percent smaller than $i_{\text{max}} = 19.5$ case.
Although $i_{\text{max}} = 21$ observation provides $3 \-- 4$ times larger number of observed stars, 
the improvement of the $J$-factor estimation is rather mild.
This is because the error below 
$\delta \log J \sim 0.1$ is dominated by the degeneracy of the parameter $r_s$, $\rho_s$, $\gamma$, and $\beta_{ani}$, 
which is hard to reduce by the projected kinematical data, as one can see in \citet{Bonnivard:2014kza}. 
This degeneracy will be resolved by measuring the proper motion of the stars by the future photometric long-time observations.
By the same reason, 
the four-pointing case and $i_{\text{max}} = 21.5$ case do not significantly improve the $J$-factor estimation, although the error of the $J$-factor becomes slightly small in both cases. It implies in turn that only the one-pointing observation provides enough stellar data to construct both member and foreground distribution.

%Since only the part of the member star can be observed, 
%the observed star may have some bias to the distribution.
The second uncertainty in the table shows the sampling fluctuation
coming from the quality of the data,
which is not included in the uncertainty of the conventional kinematical fit.
%is subdominant but not negligible.
%The most conservative estimation can be derived by including 
%The result provides the uncertainty 
%coming from the quality of the data, 
The result shows that this fluctuation is subdominant but not negligible
especially for the $i_{\text{max}} = 19.5$ case,
in which the median value of the $J$-factor has $\Order{0.1}$ fluctuation.
As for the number of the samples becomes about $3 \-- 4$ times larger, 
the $i_{\text{max}} = 21, \, 21.5$ cases provide about half of the sampling error of $i_{\text{max}} = 19.5$ case.
The difference between $i_{\text{max}} = 21$ and $i_{\text{max}} = 21.5$ cases is not obvious 
since the number of the observed stars is similar.

As an example of the result, we show the dispersion curve obtained from the best fit parameter and mock data in Fig~\ref{fig:disp}. Although we do not adopt the binned analysis, the fit successfully reproduces the input curves.
The `{\sl Contaminated}' column, on the other hand, shows that even if the contamination is 3-10 percent (see Table~\ref{tb:Cut}) and the number of the observed stars is large, the fit gives significantly large $J$-factor ($\delta \log J \sim$ 0.3-0.5). This systematic error stems from the enhancement of the velocity dispersion by the foreground stars at the outer region. 

Although our likelihood estimation gives consistent results with the `{\sl Input}' values, small discrepancies appear for the Draco\,1  and Ursa\,Minor case. In these fits, the likelihood distribution has a broad and flat peak (especially along the $r_s$ axis) 
and the scattered points around the input parameter contribute to the $J$-factor distribution asymmetrically around the input $J$-factor value.
As a result, the distribution of the $J$-factor is distorted from a Gaussian shape
and its peak does not always coincide with the input $J$-factor value,
which generates the discrepancy between the median and input $J$-factors.
The tendency of this $J$-factor bias is basically determined by some combinations of the input parameters and the precise prediction is rather difficult. 
For example, for Ursa\,Minor case, the results give slightly higher $J$-factors than the input value. 
We have checked this tendency by using the member star only data and found that for $i_\text{max} = 21, \theta_{\text{ROI}} = 0.65 \unit{degree}$, the fit gives $\log_{10} (J/[\text{GeV}^2/\text{cm}^5]) = 19.09^{+0.13}_{-0.08} \pm 0.03$ for Ursa\,Minor. 
For the Draco\,1 case, on the other hand, the fit using only the member stars gives $\log_{10} (J/[\text{GeV}^2/\text{cm}^5]) = 18.86^{+0.09}_{-0.08} \pm 0.08$, which is slightly smaller than the input $J$-factor.
These results imply that even if we could perfectly reduce the foreground star, 
the degeneracy of the parameter gives a small discrepancy of the $J$-factors. 

For the Draco\,1 case, 
the foreground contamination gives another bias in our fit, 
allowing smaller $r_s$ region under a fixed halo radius $\rho_s$.
This bias can be explained as follows: Consider the stellar distribution with respect to its velocity (like Fig.\,\ref{fig:FGdist}). 
In the region of $v_{\text{lower}} < v < v_{\text{upper}}$, the foreground velocity distribution monotonically increases as the velocity increases. 
On the other hand, the member stars distribute like Gaussian at $v = v_{\text{dSph}}$ with a standard deviation of $\sigma_{l.o.s} (R)$. 
The properties of these distributions imply that more miss-identification occur in the $v > v_{\text{dSph}}$ region, where the number of the foreground stars increases. 
Due to this asymmetric uncertainty, the fit tends to underestimate the width of the velocity distribution of the member stars. 
In particular, when the dark matter halo size $r_s$ is large like the Draco\,1 case, the velocity dispersion of the member stars becomes large at the outer region $R \gtrsim \Order{1} \,\text{kpc}$, in which the fraction of the foreground star increases. 
As a result, the identification becomes more difficult in this outer region, deriving the underestimation of the halo size as in the case of Draco\,1.
One should, therefore, be aware of the systematic errors when the resultant $r_s$ is, roughly speaking, larger than the maximum distance of the observed stars.

The same approach can be applied to the other classical dSphs.
For larger dSphs such as Fornax, Sculptor, and Carina, observations with multiple-pointings may be required to cover all the member stars. 
However, even in that case, 
the total exposure time does not inflate because the number of the bright member stars are sufficiently large.
Moreover, the multiple-pointings observation will also bring the information about the tidal radius of each dSph.

Before closing this section, we stress that the result of the current data (corresponding to the contaminated case of $i_{\text{max}} = 19.5$) indicates the existence of a non-negligible systematic bias from the foreground stars even for the classical dSph with \Order{100} stellar data. Usually, the foreground stars are eliminated by imposing the membership probability\,\citep{Walker:2008fc},
in which the stellar data with the probability above 95\,\% is considered as a member star. However, our analysis shows that only 5\,\% contamination significantly leads to the overestimation of the $J$-factor by a factor of three
and gives an incorrectly stringent limit to the dark matter annihilation cross section.
Therefore, we conclude that a careful treatment of the foreground estimation is required to study the dark matter nature.

%------------------------------------------------------------------------------
%-----------------------          Summary          ----------------------------
%------------------------------------------------------------------------------
\section{Summary}
\label{sec:summary}
In this paper, we have investigated the future impact of the stellar observation and the effect of the foreground contamination by using the mock dSph samples.
We have tested the various cuts to optimize the quality of the data and found that the cuts on the velocity and surface gravity efficiently eliminate the contamination, while other cuts do not work well because the member stars and the halo stars degenerate.
A new likelihood function has been constructed which includes the foreground distribution function.
We have tested the likelihood function by making the three types of the mock data (Draco with large $r_s$, Draco with small $r_s$, and Ursa\,Minor) and four cases of the observation (small/large ROI, $i_{\text{max}}=$19.5, 21, and 21.5).
The likelihood successfully reproduces the input $J$-factor value while the contaminated fit gives large deviation from the input value.
The small discrepancy between the input $J$-factor and median value obtained by our fit 
suggests that 
the parameter degeneracy in the likelihood function distorts the $J$-factor distribution 
from the Gaussian shape with a peak at the input value and gives an asymmetric distribution.
This non-Gaussianity may affect the estimation of the sensitivity lines of dark matter detection.
We have also found that the reduction of the foreground effect becomes worse when the halo radius is large (roughly larger than the outermost observed star), which causes  $\delta \log J \sim 0.2$ deviation.
The effect of the sampling fluctuation has also been estimated. It is found that the statistical fluctuation of the sampling leads $\delta \log J \sim 0.1$ at most even for the fit under ${\cal O}(1000)$ samples.

%---------------------------------------------------------------------
%---------------------          Acknowledge          --------------------------
%---------------------------------------------------------------------
\vspace{0.5cm}
\noindent
{\bf Acknowledgments}
\vspace{0.1cm}\\
\noindent
This work is supported in part by a Grants-in-Aid for Scientific Research from the Japan Society for the Promotion of Science (JSPS) (No. 25-7047 for M. N. I. and No. 26-3302 for K.H.) and
from the Ministry of Education, Culture, Sports, Science, and Technology (MEXT), Japan, 
No. 16H01090 (for K. H.),
No. 15H05889, No. 25105011 (for M. I.), 
No. 16H02176, No. 26104009 (for S. M.), 
and 
No. 26287039 (for S. M. and M. I.).
The work of K.I. is also supported by the JSPS Research Fellowships for Young Scientists.
Finally, Kavli IPMU is supported by World Premier International Research Center Initiative (WPI), MEXT, Japan.

%%%%%%%%%%%%%%%%%%%%%%%%%%%%%%%%%%%%%%%%%%%%%%%%%%


\bibliographystyle{mnras}
\bibliography{ref2}

%%%%%%%%%%%%%%%%%%%%%%%%%%%%%%%%%%%%%%%%%%%%%%%%%%

%%%%%%%%%%%%%%%%% APPENDICES %%%%%%%%%%%%%%%%%%%%%

\appendix

\section{Foreground Distribution}
\label{app:FG}

In this appendix, we mention the method to obtain the foreground prior discussed in Sec.\,\ref{subsec:pfslikelihood}.
The foreground stars are mainly composed of three components: the halo stars, the thick disc component, and the thin disc component, for the line-of-sights of both Draco and Ursa\,Minor.
To determine the prior, we use $v$ and $\log g$ information of the mock observation data.
For the explanation, we categorize the dataset after the colour and ROI cuts into two types.
\footnote{
The number of the stars (including the member and foreground stars) after the colour and ROI cuts is $\sim 2000$ for the one-pointing case.
Therefore, all the stellar data  utilized in this analysis can be covered by the PFS observation (with 2394 fibres).
}
\begin{enumerate}
\item the dataset with velocity cut.
\item the dataset with velocity \& $\log g$ cut.
\end{enumerate}
We note that the velocity cut here implies masking the signal region ($v_{\text{lower}} < v < v_{\text{upper}}$) to obtain the pure foreground samples.
The goal is to determine the foreground shape in the dataset\,(ii).
In the dataset\,(ii), although the dominant contribution is the halo component, the other components non-negligibly distort the shape of the foreground distribution.
Therefore, we fit all components assuming that their velocity distribution can all be described by the Gaussian distribution function.
However, because the disc distributions are located at around $\log g \gtrsim 4$, the number of the disc components after the $\log g$ cut becomes so small that the fit cannot converge well.
Therefore, we first determine the shape of the thick/thin disc component using the data without $\log g$ cut (dataset\,(i)).

\begin{figure}
\begin{center}
\includegraphics[scale=0.37]{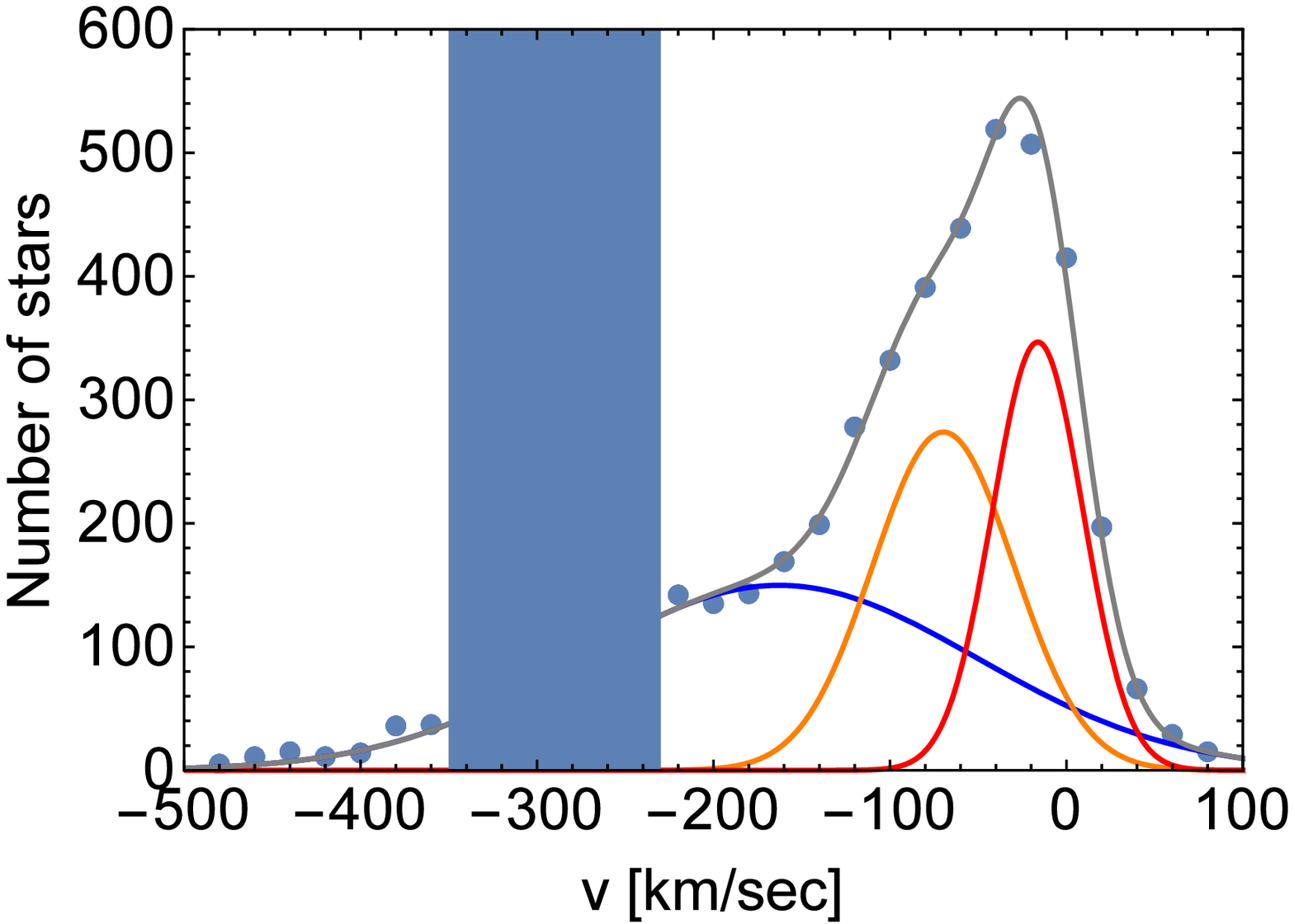}
\includegraphics[scale=0.36]{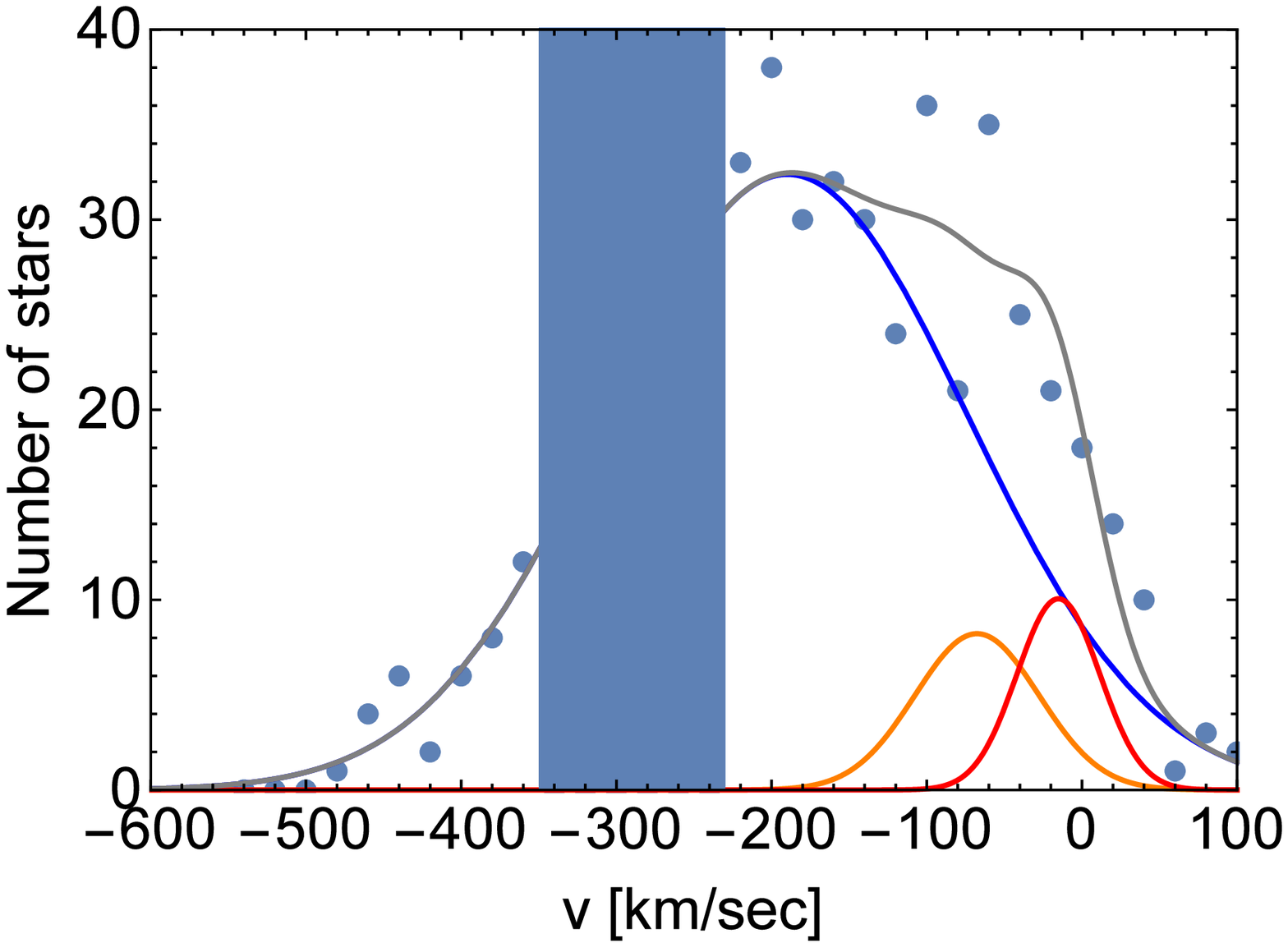}
\caption{\small\sl Foreground fit for the Draco\,1 sample with $\theta_{\text{ROI}} = 1.3,\,i_{\text{max}} = 21$.
The blue-shaded region is the signal region.
The grey line shows the result of the fit and the blue, orange, red lines show the contribution from the halo, thick disc, 
and thin disc component respectively.
The top panel is the result of the dataset\,(i) while the bottom is of the dataset\,(ii). See the text for more detail.
}
\label{fig:FGdist}
\end{center}
\end{figure}

In the first fit, we fit the velocity distribution of the dataset\,(i) by the sum of the three Gaussians allowing all the parameters  (normalizations, mean velocities, dispersions) free.
As one can see in the Fig.\,\ref{fig:FGdist}, the fit can be successfully performed because the peaks of the three components are obvious.
Then, assuming that the distributions of the thick/thin disc component do not change after the $\log g$ cut (except for their normalization), we use these mean velocities and dispersions to the next fit (dataset\,(ii)).
\footnote{We have checked the $\log g$ dependence on these components and found that it is small.}
Here we note that we cannot utilize the information of the halo distribution to the 
second fit
because the halo distribution non-negligibly depends on the $\log g$.\footnote{
The shapes of the halo component keep Gaussian in the range of interest, while its mean and width change with respect the $\log g$ value.
}

In the second fit, we again assume that the three components are the normal distribution and fit them to the dataset\,(ii) (with velocity and $\log g$ cut).
In this fit, we constrain the shapes of the disc components by imposing the Gaussian prior of the mean velocities and the dispersions, which are obtained from the first fit.
The example of the fit is shown in the bottom panel of Fig.\,\ref{fig:FGdist}.

We finally note that the foreground curve in the signal region (the blue-shaded region in Fig.\,\ref{fig:FGdist}) is dominated by the halo component and therefore the prior function for the main fit in Sec.\,\ref{subsec:pfslikelihood} can be described by the single Gaussian of the halo component.

\section{Comparison to the conventional method}
\label{app:EM}

In Sec.\,\ref{sec:result}, we compare the results of our $J$-factor estimation with the `{\sl Contaminated fit}' analysis 
where we regard the foreground stars after the naive cuts in Sec.\,\ref{subsec:cut} as the member stars.
%In Sec.\,\ref{sec:result}, we compare the results of our $J$-factor estimation with 
%the `Contaminated' fit by using the data imposed the naive cuts in Sec.\,\ref{subsec:cut}.
In the conventional analysis in \citet{Geringer-Sameth:2014yza}, 
the foreground stars are also treated as the member stars 
as in our `{\sl Contaminated fit}' to estimate the $J$-factor, 
although more sophisticated data cuts are 
applied so that the fraction of the member stars in the data set increases. 
In this section, we compare the results obtained by 
the conventional analysis (sophisticated extraction scheme + naive contaminated fit) with ours (naive cut + our likelihood fit).
%In the conventional analysis in \citealp{Geringer-Sameth:2014yza}, 
%though the fit process is the same as the `Contaminated' one,
%more sophisticated data cut scheme is adopted to
%increase the fraction of the member stars in the data\,\citep{Walker:2008fc}.
%The distributions of the foreground  and member stars after this scheme
%can differ from the naive data cut which we adopt 
%and therefore it is interesting to compare the results obtained by 
%the conventional method (sophisticated extraction scheme + naive contaminated fit) 
%with ours (naive cut + our likelihood fit).

%Below, we first review the calculation process of the membership probability, giving the relation to our fit
%and using the mock data obtained by Sec.\,\ref{subsec:mock},
%we provide the results of the conventional method.

Below, we first review the calculation process of the membership probability, giving the relation to our fit. 
We specifically consider the method adopted in \,\citet{Walker:2008fc} based on 
the Expectation Maximization (EM) algorithm. 
We then provide the results of the conventional method using the mock data obtained in Sec.\,\ref{subsec:mock}.

\subsection{Basics}

In contrast with our analysis, 
the conventional analysis assigns a membership probability $\emp(X_i)$ to each stars.
Here $X_i$ denotes the information of $i$-th star  (position, velocity, metalicity, etcetera).
The $J$-factor is estimated by regarding all the stars with a given confidence level of 
$\emp(X_i)$, (e.g. $\emp(X_i) > 0.95$), as member stars.
Therefore, in the conventional way the robust estimation of $\emp(X_i)$ plays important role 
for the $J$-factor estimation.

If we could distinguish the members from the foregrounds,  
the membership probability would be given by the probability to find a member star at the specified distance, such as
\begin{eqnarray}
	\emp(X_i) \equiv p(M_i=1|X_i,\theta_\tMem, \theta_\tFG, s)\ ,
\end{eqnarray}
where the new observable $M_i$ takes a boolean value 
which indicates whether $i$-th star is a member ($M=1$) or a foreground ($M=0$), 
introduced for the convenience. ~\footnote{
	We use notation $p(X|Y)$ for the posterior probability (or probability density function, p.d.f.) of $X$
	after the observation of $Y$.
}
The parameters $\theta_\tMem$, $\theta_\tFG$ and $s$ are the free parameters 
describing the member and foreground distributions.

The explicit form of $\emp(X_i)$ can be derived from 
the fundamental likelihood
$\Likeli_0\equiv \prod_i p(M_i,X_i|\theta_\tMem, \theta_\tFG, s)$ 
where we define
\begin{eqnarray}
	\ln \Likeli_0 =
		\sum_i {M_i}\ln(sf_\tMem(X_i)) \hspace{80pt} \nonumber \\
		\hspace{50pt} + (1-M_i)\ln((1-s)f_\tFG(X_i))\ .
\label{eq:L0}
\end{eqnarray}
The explicit form of $\emp(X_i)$ is given by
	\begin{eqnarray}
		\emp(X_i)=
			\frac{\displaystyle s \emfmem(X_i)}
				{\displaystyle s \emfmem(X_i) +(1-s)\emffg(X_i)}\ .
	\end{eqnarray}
We note that our likelihood $\Likeli$ is equal to 
$p(X_i = \{v_i,\,R_i\}|\theta_\tMem, \theta_\tFG, s) = \prod_i\sum_{M_i=1,\,0}\Likeli_0$.

In the conventional analysis, 
likelihood method is used to maximize 
$\prod_ip(X_{i/R_i}|R_i,\theta_\tMem,\theta_\tFG,s)$ 
and the parameter at the maximum is used to evaluate $\emp(X_i)$.
Here $X_{i/R_i} $ represents the stellar information except for its position.
For this maximum search, 
Expectation Maximization (EM) algorithm is adopted,
which enables us to estimate $\emp(X_i)$ 
incidentally through the recursive calculation based on 
$\emLikeli \equiv \prod_i p(M_i,X_{i/R_i}|R_i, \theta_\tMem, \theta_\tFG, s)$
 instead of the direct maximization of its integration 
 $\prod_i p(X_{i/R_i}|R_i,\theta_\tMem,\theta_\tFG,s) = \prod_i \sum_{M_i}\emLikeli$.
Here, by using the expression in Eq.(\ref{eq:L0}), the likelihood $\emLikeli$
can be written as 
	\begin{eqnarray}
		\displaystyle\ln\emLikeli\ \equiv
			\sum_i {M_i}\ln \left(\frac{\emfmem(X_i)}{\emfmem(R_i)}t(R_i)\right) 
			\hspace{60pt} \nonumber \\
		\hspace{50pt} + (1-M_i)\ln \left(\frac{\emffg(X_i)}{\emffg(R_i)}(1-t(R_i))\right).
	\label{eq:EMlikeli}
	\end{eqnarray}
Here we introduce $\emfmem(R_i)\equiv\int dX_{i/R_i}\,\emfmem(X_i)$ and $\emffg(R_i)\equiv\int dX_{i/R_i}\,\emffg(X_i)$, 
for the convenience.
$t(R_i) \equiv s \emfmem(R_i)/(s \emfmem(R_i) +(1-s)\emffg(R_i)) $ represents the member fraction 
of the stellar spatial distribution.
Note that $\emfmem(X)/\emfmem(R)$ and 
$\emffg(X)/\emffg(R)$
corresponds to 
the velocity distribution function 
${\cal G}[v;\,v_{\text{Mem}},\,\sigma_{l.o.s} (R)]$
and  ${\cal G}[v;\,v_{\text{FG}},\,\sigma_{\text{FG}}]$
of our analysis respectively.

In addition, the models of the distribution functions used in the conventional method are different from ours.
First, no specific functional form of the spatial distribution functions 
$\emfmem(R)$, $\emffg(R)$ are assumed,
while we adopt the Plummer profile and linear increasing function of $R$ respectively.
Rather, they only assume that $t(R_i)$ is monotonically decreasing function of $R$
and optimize it considering the value of  $t(R_i)$ at each $R_i$ as a free parameter.
Second, they use the velocity distribution functions defined by
	\begin{eqnarray}
		\frac{\emfmem(v,R)}{\emfmem(R)} &=& {\cal G}[v;\,\emvmem,\,\emsmem]\ ,\\
		\frac{\emffg(v,R)}{\emffg(R)}  &=& \frac{1}{N_\tBes}\sum_{i=1}^{N_\tBes} {\cal G}[v;\,\emv_{\tFG_i},\,\emsfg]\ ,
	\end{eqnarray}
where $\emvmem$ and $\emsmem$ are free parameters of the likelihood function $\emLikeli$,
whereas $\emv_{\tFG_i}$,  $\emsfg$ and $N_\tBes$ are fixed parameters 
achieved by the numerical simulation based on the \Besancon\ model.
We here stress that 
this assumption means that $\emsmem$ does not depend on $R$ 
and therefore 
the membership probability obtained by this process is biased 
to have a constant velocity dispersion curve.
Moreover, in the conventional analysis, 
$\emffg(v)$ is completely fixed by the numerical model
while our distribution function has free parameters $v_\tFG,\,\sigma_\tFG$ in the Gaussian.
We also note that the conventional analysis uses
stars with a membership probability of a given confident level for the $J$-factor estimation.
In this fit the systematic bias caused by the foreground contamination is not taken into account.
This effect could be included in the fit by modifying 
the likelihood function given by Eq.\,(\ref{eq:fmem})
introducing weights from the membership probability for each star\,\citep{Bonnivard:2015vua}.
However, since the EM method only provides the parameter at the maximum of the likelihood,
one cannot obtain the error of each membership probability.
Therefore, the systematic uncertainty from this error is not included 
even in the modified fit.
In our analysis, in contrast,
we do not need to handle this systematic uncertainty because it is automatically included 
in the error bars obtained by the MCMC sampling method using our likelihood function.

\subsection{$J$-factor estimation using mock data}

\begin{table}
  \begin{tabular}{ccccc}
  dSph
& $\theta_{\text{ROI}}$ [degree] 
& $i_{\text{max}}$ [mag] 
& $N_{\text{Mem}}$
& $N_{\text{FG}}$
\\ \hline
Draco 1 \& 2 & $0.65$ & $19.5$  & 270 &  8 \\
             &        & $21$    &     900 &  11 \\
             &        & $21.5$  &    1140 &  12 \\
             & $1.3$  & $21$    &   920 & 22 \\ \hline
Ursa\,Minor  & $0.65$ & $19.5$  &   290 &  3 \\ 
             &        & $21$    &            1120 &  11 \\ 
             &        & $21.5$  &            1450 &  14 \\ 
             & $1.3$  & $21$    &         1140 & 16 \\ \hline
  \end{tabular}
\caption{\sl
The number of the member (foreground) stars with membership probability above $95\,\%$
are given by $N_{\text{Mem}}$ ($N_{\text{FG}}$). 
50 mocks are generated by the same procedure as Sec.\,\ref{subsec:mock} and
the membership probability for each star is calculated by the EM method (see the text and 
\citet{Walker:2008fc} for the details).
}
\label{tb:EM}
\end{table}

In this section, we give the results of the $J$-factor estimation by the conventional fit.
We generate the mock data by the same process in Sec.\,\ref{subsec:mock}.
Here we do not impose any cuts to the data except for the color magnitude cut.
Instead, the EM method given by \citet{Walker:2008fc} is applied to each mock
using the information of  [Fe/H], $\log g$, $v$ and $R$.
The membership probability is assigned to each star after through the EM algorithm
and the stars with the membership probability above $95\,\%$ are extracted.
Table~\ref{tb:EM} shows the number of the member and foreground stars after this process (averaged by the 50 mocks).
The fraction of the foreground stars is less than $5\,\%$ 
and better than those obtained by the naive cut process in Sec.\,\ref{subsec:cut}
(see Table~\ref{tb:Cut}).
This is because
the information of the spacial distribution is used in the EM method,
which is not involved in the cut process.
%(rather, in our method, it is used in the optimization of the likelihood given by the Eq.(\ref{eq:likeli}).
Moreover, in the EM method, 
more accurate discrimination can be achieved
by the weights from the [Fe/H] and $\log g$ 
through the optimization of the means and widths of these Gaussian distributions.

\begin{figure}
 \begin{center}
  \includegraphics[width=80mm,angle=0]{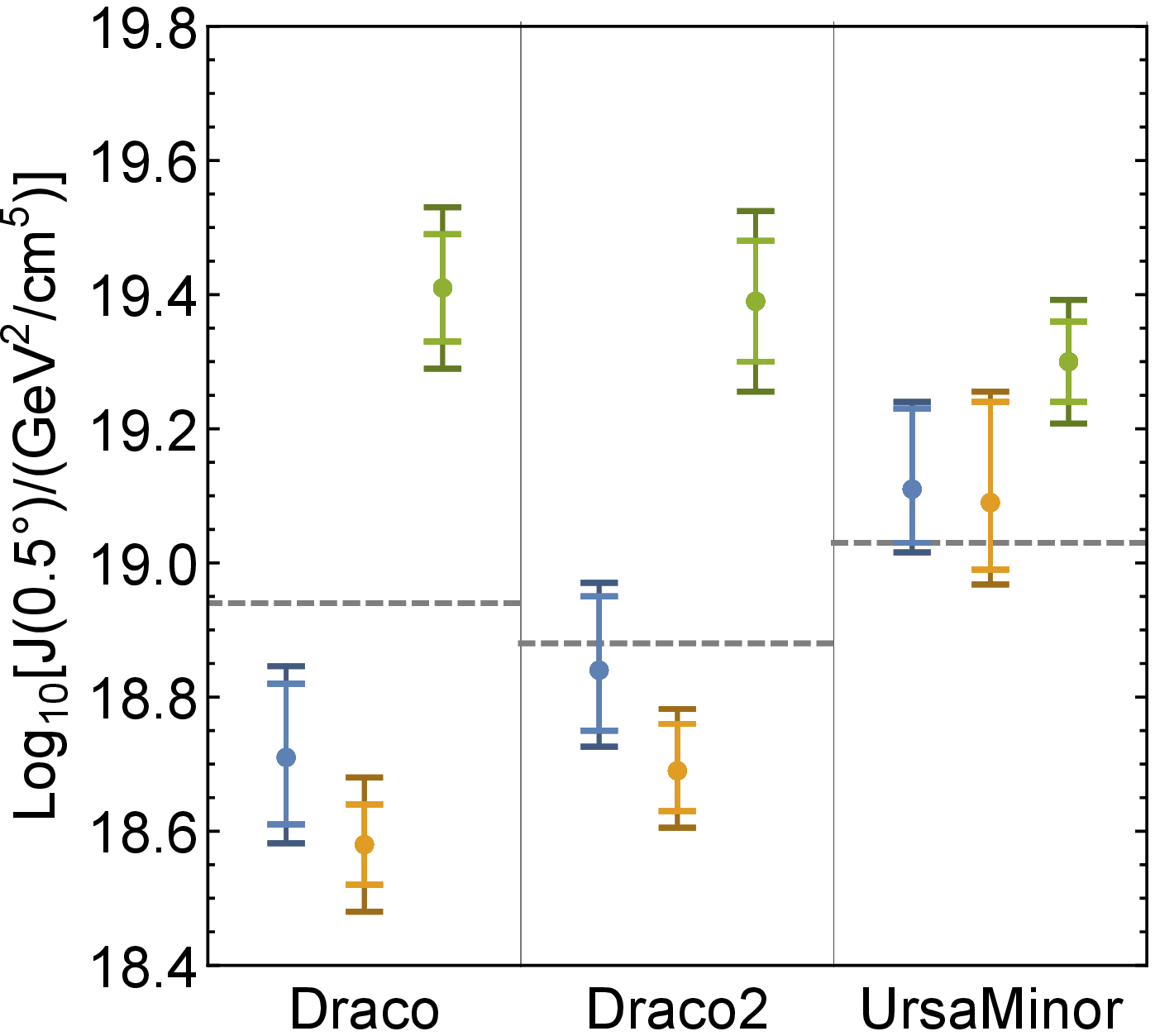}
\caption{\small\sl
The $J$-factors obtained by the conventional method for 
the case of $i_{\text{max}} = 21$ with $\theta_{\text{ROI}} = 0.65$
are plotted (orange bars). 
For comparison, 
$J$-factor estimation of `{\sl Our fit}' (blue bars) and  `{\sl Contaminated}' (green bars) are also shown. 
The lighter error bars show the average of the $68\,\%$ quantile, 
while the darker ones show the square root of the $68\,\%$ quantiles and the standard deviation of the median values.
}
 \label{fig:JEM}
 \end{center}
\end{figure}

\begin{figure}
 \begin{center}
  \includegraphics[width=70mm,angle=0]{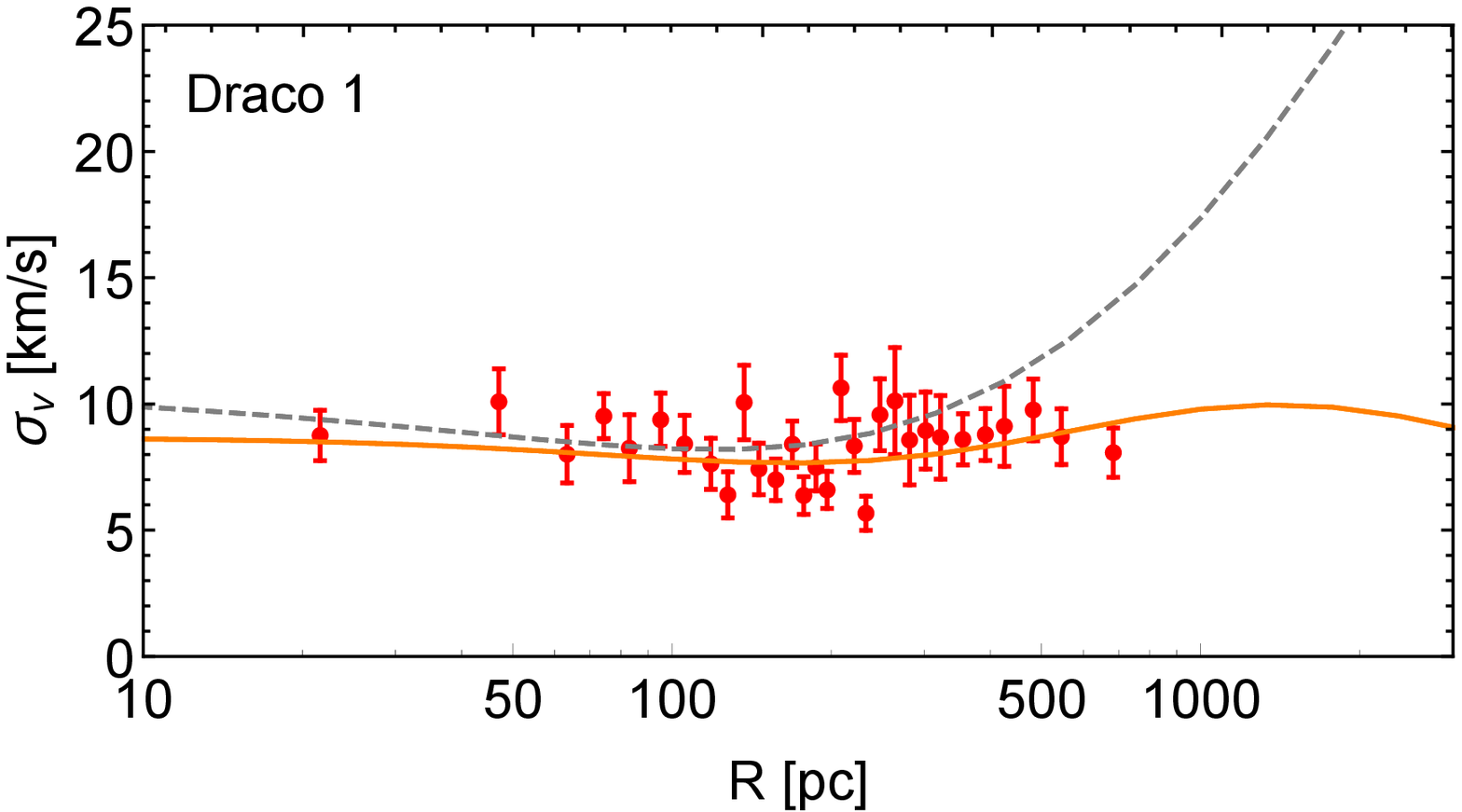}
\\
  \includegraphics[width=70mm,angle=0]{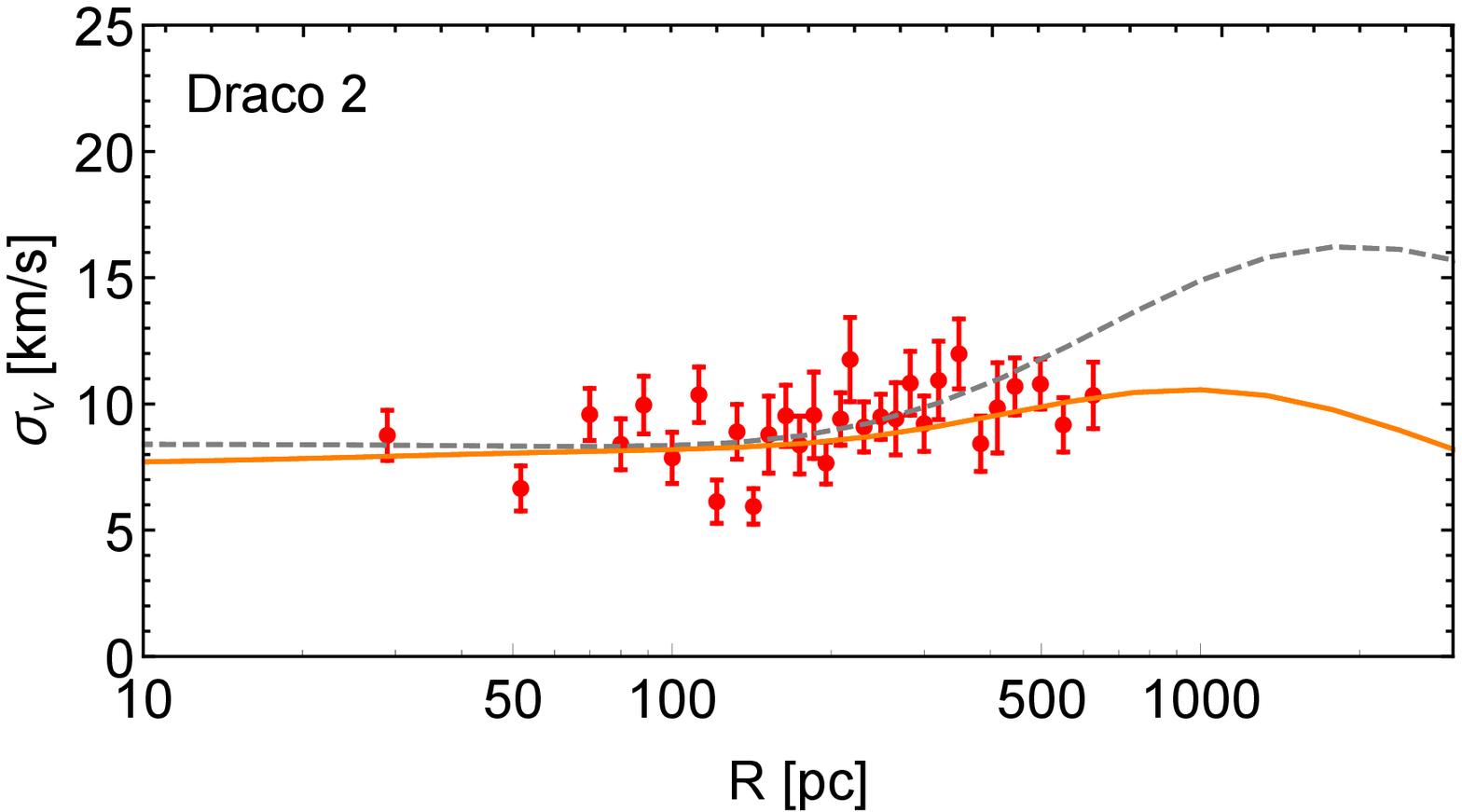}
\\
  \includegraphics[width=70mm,angle=0]{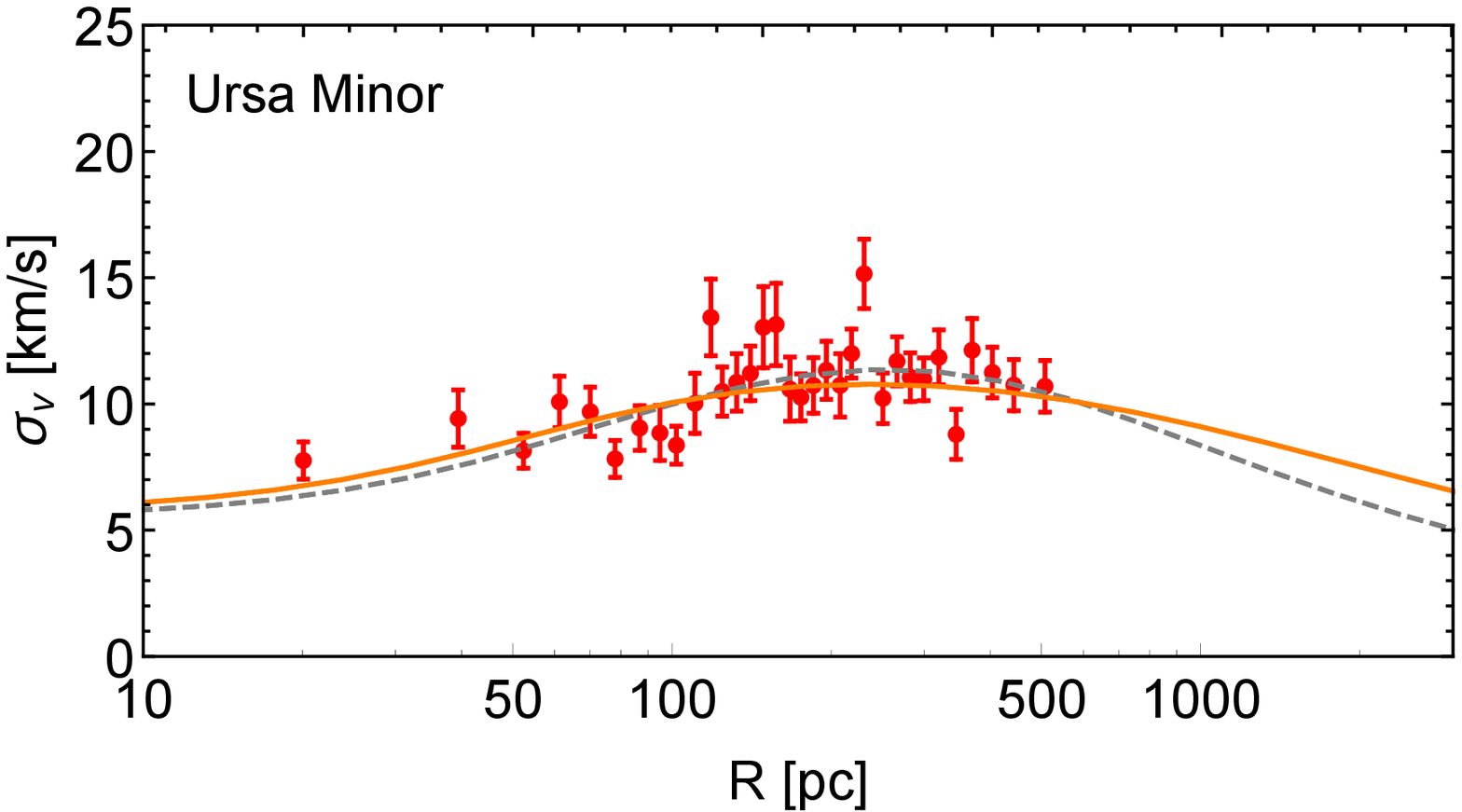}
 \end{center}
 \caption{\small\sl
The dispersion curve of $\theta_{\text{ROI}} = 0.65,\,i_{\text{max}} = 21$ case of 
Draco\,1 (top), Draco\,2 (middle), and Ursa\,Minor (bottom).
The binned dispersions of the mock data after the EM selection 
are shown by the red dots with error bars.
The orange line shows the dispersion curve of the best fit.
The dashed line is obtained from the input parameter of the dSph dark matter halo.
}
 \label{fig:EMdisp}
\end{figure}

Using this extracted data, 
we estimate the $J$-factor by the same method as `{\sl Contaminated}' case
(in which, i.e., the data is fitted by using $f_{\text{Mem}}$ in Eq.(\ref{eq:fmem}) 
accepting all the extracted star as the member star).
%therefore it is interesting to compare the $J$-factors obtained by the convensional process  
%with those obtained by `{\sl Our fit} and `{\sl Contaminated}' fit defined in Sec.\,\ref{sec:result}.
We provide the result of the case of  $i_{\text{max}} = 21$ with $\theta_{\text{ROI}} = 0.65$ 
in Fig.\,\ref{fig:JEM}.
The figure shows that even though the contamination is less than $5\,\%$,
mild systematic biases appear in the conventional method for Draco\,1 and Draco\,2 cases.
This systematic bias decreases the median value of the $J$-factor,
in contrast to the overestimation of the `{\sl Contaminated}' fit.
This is because the distribution of the stars is different from the `{\sl Contaminated}' data.
In the conventional approach, the constant velocity dispersion is assumed 
and hence the extracted data can be  biased to have the constant dispersion.
This tendency is obvious for the two Draco cases, 
where the input dispersion curves largely increase at the outer region
as shown in Fig.\,\ref{fig:EMdisp}.
We stress that 
this bias does not appear in our process because 
we adopt the range of the naive cuts wide enough
in order not to distort the velocity distribution.\footnote{
In the cut process, 
some fraction of the member stars is eliminated from the raw data
by the naive $\log g$ cut.
However, since $\log g$ cut is independent from the $r$, $v$ distribution of the member stars,
this naive cut does not affect the dispersion curve.
}
%the higher membership probability is assigned to the stars at around the centric region,
%and therefore the foreground contamination tends to concentrate at the inner region of the dSphs,
%which might generate systematical error.
%Since we only test the three types of the dSphs, 
%we can not draw a conclusion for this bias.

Before closing this appendix, we comment on the ultra-faint dSphs case.
Since the number of the member stars is much smaller than the classicals,
the foreground contamination can be large even after the EM selection.
Moreover, as discussed above, 
the conventional method has the bias of the constant velocity dispersion.
In the ultra-faint dSphs, the parameter of the constant dispersion 
is controlled by the large foreground contamination and small member stars
 and therefore can often be misidentified.
% which often eliminates some member stars to maintain a constant dispersion curve.
These effects can be a dominant hidden systematic uncertainties for the ultra-faint dSphs case,
deriving a huge deviation from the input $J$-factor.
%As shown in this paper,
Due to the simultaneous fit of the member and foreground distribution
including the $R$ dependence of the dispersion curve, 
our analysis will provide safer estimation especially for the ultra-faint dSphs,
as demonstrated for classical dSphs in this paper.
%as shown in this paper.
We will give detailed results for the ultra-faint dSphs case in the forthcoming work\,\citep{Ichikawa:ref}.

%Although the detailed origin  of this bias is not clear,
%it might be because the distribution of the foreground stars are different from the `{\sl Contaminated}' data.
%In the EM procedure, the constant velosity dispersion is assumed 
%and therefore the extracted data can be biased to have the constant dispersion.
%the higher membership probability is assigned to the stars at around the centric region,
%and therefore the foreground contamination tends to concentrate at the inner region of the dSphs,
%which might generate systematical error.
%Since we only test the three types of the dSphs, 
%we can not draw a conclusion for this bias.
%We will provide a detailed analysis in the forthcoming work of ultra-faint dSphs\,\cite{xxx}, 
%in which we expect that more obvious effect of the contamination can be seen.

\section{Effect of the truncation radius}
\label{app:RTrunc}

\begin{figure}
 \begin{center}
  \includegraphics[width=80mm,angle=0]{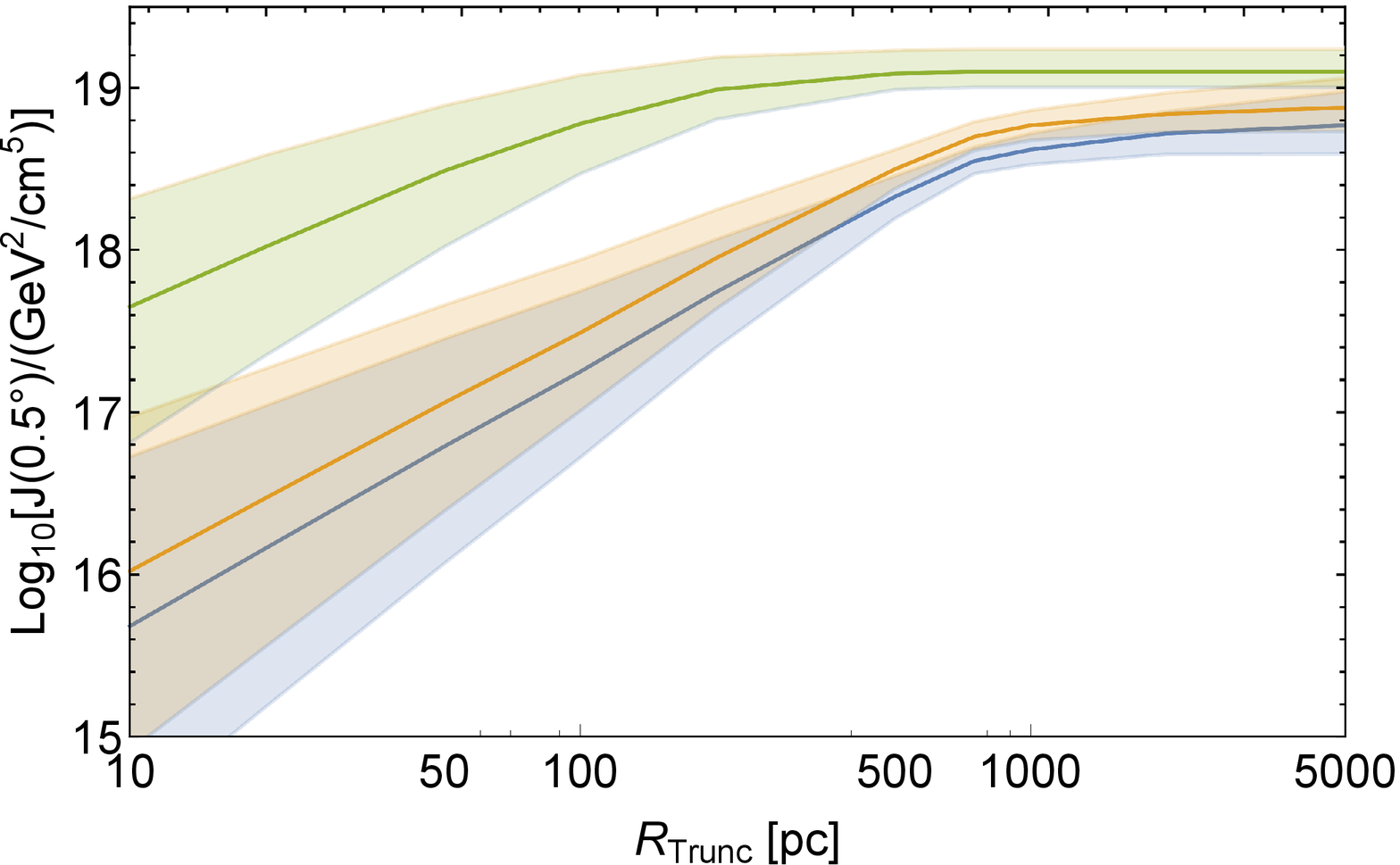}
\caption{\small\sl
The dependence of the truncation radius for the $J$-factors for 
the case of $i_{\text{max}} = 21$ with $\theta_{\text{ROI}} = 0.65$.
The blue, orange and green lines show the $J$-factor median values of 50 mocks  
of  Draco\,1, Draco\,2, and Ursa\,Minor  respectively.
The lighter bands show the averages of the $68\,\%$ quantiles.
}
 \label{fig:JRTrunc}
 \end{center}
\end{figure}

In our analysis, we set the truncation radius of the dark matter halo to $2000$ pc.
By using the MCMC samples obtained by our fit,
we check the dependence of the truncation radius for the $J$-factor and the size of the error bars.
Fig.\,\ref{fig:JRTrunc} shows this dependence
and we find that 
the fluctuation of the $J$-factor error is not largely affected by 
the truncation radius when it is above $1000$ pc
and the error is minimized at around $800$ pc.

%%%%%%%%%%%%%%%%%%%%%%%%%%%%%%%%%%%%%%%%%%%%%%%%%%

% Don't change these lines
\bsp	% typesetting comment
\label{lastpage}
\end{document}